\documentclass[journal,10pt]{IEEEtran}

\usepackage[dvipsnames]{xcolor}  
\usepackage{subcaption}
\usepackage{tikz,caption}
\usetikzlibrary{calc}
\DeclareCaptionFormat{overlay}{\gdef\capoverlay{#1#2#3\par}}
\DeclareCaptionStyle{overlay}{format=overlay}

\usepackage{amsmath, amsthm, amssymb, mathrsfs}
\usepackage{tikzsymbols} 

\usepackage[colorlinks=true, linkcolor = blue,
            urlcolor  = blue,
            citecolor = blue,
            anchorcolor = blue]{hyperref}
\usepackage{url} 

\usepackage{lipsum} 

\usepackage{multirow}
\usepackage{multicol}
\usepackage{booktabs}
\usepackage[para,online,flushleft]{threeparttable}

\usepackage{authblk}

\graphicspath{ {figs/} } 
\usepackage{float}

\usepackage[numbers,sort&compress]{natbib} 

\usepackage{wrapfig}
\usepackage{enumitem} 

\usepackage{natbib}
\newcommand{\smalltilde}{\raise.17ex\hbox{$\scriptstyle\mathtt{\sim}$}} 

\usepackage[ruled,vlined]{algorithm2e}
\SetAlFnt{\small}
\SetAlCapFnt{\small}
\SetAlCapNameFnt{\small}

\usepackage{caption}
\usepackage{subcaption}
\usepackage{xargs}                      

\usepackage[colorinlistoftodos,prependcaption, textsize=small]{todonotes}
\newcommandx{\unsure}[2][1=]{\todo[disable, inline,linecolor=red,backgroundcolor=red!25,bordercolor=red,#1]{#2}}
\newcommandx{\add}[2][1=]{\todo[disable, inline,linecolor=green,backgroundcolor=green!25,bordercolor=green,#1]{#2}}
\newcommandx{\fix}[2][1=]{\todo[disable, inline,linecolor=blue,backgroundcolor=blue!25,bordercolor=blue,#1]{#2}}

\usepackage{enumitem} 

  {\begin{enumerate}%
    leftmargin = *
    \setlength{\itemsep}{0pt}%
    \setlength{\parskip}{0pt}}%
  {\end{enumerate}}

\newtheorem{example}{Example} 

\newtheorem{prop}{Proposition}
\newtheorem{thm}{Theorem}
\newtheorem{cor}{Corollary}

\newtheorem{defn}{Definition}
\newtheorem{rmk}{Remark}

\usepackage{wasysym}

\usepackage{arydshln}

\makeatletter
\def\adl@drawiv#1#2#3{%
        \hskip.5\tabcolsep
        \xleaders#3{#2.5\@tempdimb #1{1}#2.5\@tempdimb}%
                #2\z@ plus1fil minus1fil\relax
        \hskip.5\tabcolsep}
\newcommand{\cdashlinelr}[1]{%
  \noalign{\vskip\aboverulesep
           \global\let\@dashdrawstore\adl@draw
           \global\let\adl@draw\adl@drawiv}
  \cdashline{#1}
  \noalign{\global\let\adl@draw\@dashdrawstore
           \vskip\belowrulesep}}
\makeatother

\definecolor{brightgreen}{rgb}{0.4, 1.0, 0.0}

\usepackage{wasysym}

\usepackage{adjustbox}
\usepackage{array}

\newcommand*{\full}{\CIRCLE}
\newcommand*{\prt}{\LEFTcircle}
\newcommand*{\none}{\Circle}

\newcommand*{\setuptable}{
	\renewcommand{\arraystretch}{1.1}
	\setlength{\arrayrulewidth}{0.1em}
	\setlength\tabcolsep{3.75pt}
	\centering
}

\newcommand*{\headrow}[1]{\multicolumn{1}{c}{\adjustbox{angle=35,lap=\width-0.5em}{#1}}}

\usepackage{xhfill}

\usepackage{framed}

\newenvironment{myproof}[1][\proofname]{%
  \begin{proof}[#1]$ $\nobreak\ignorespaces
}{%
  \end{proof}
}

\usepackage{microtype}

\begin{document}




\date{}

\title{\LARGE \bf CAN-D: A Modular Four-Step Pipeline for Comprehensively Decoding Controller Area Network Data
\thanks{\small{
Copyright (c) 2015 IEEE. Personal use of this material is permitted. However, permission to use this material for any other purposes must be obtained from the IEEE by sending a request to \href{mailto:pubs-permissions@ieee.org}{pubs-permissions@ieee.org}.
This manuscript has been co-authored by UT-Battelle, LLC, under contract DE-AC05-00OR22725 with the US Department of Energy (DOE). The US government retains and the publisher, by accepting the article for publication, acknowledges that the US government retains a nonexclusive, paid-up, irrevocable, worldwide license to publish or reproduce the published form of this manuscript, or allow others to do so, for US government purposes. DOE will provide public access to these results of federally sponsored research in accordance with the DOE Public Access Plan (\url{http://energy.gov/downloads/doe-public-access-plan}).
}}}





\author{
    \IEEEauthorblockN{
    Miki E. Verma\IEEEauthorrefmark{1}, 
    Robert A. Bridges\IEEEauthorrefmark{1}, 
    Jordan J. Sosnowski\IEEEauthorrefmark{2}, 
    Samuel C. Hollifield\IEEEauthorrefmark{1}, 
    Michael D. Iannacone\IEEEauthorrefmark{1}} \\
    \IEEEauthorblockA{\IEEEauthorrefmark{1}\small Cyber \& Applied Data Analytics Division, Oak Ridge National Laboratory, Oak Ridge, TN
    \\\{vermake, bridgesra, hollifieldsc, iannaconemd\}@ornl.gov} \\
    \IEEEauthorblockA{\IEEEauthorrefmark{2}Department of Computer Science \& Software Engineering, Auburn University
    \\jjs0029@auburn.edu}
}

\markboth{IEEE Transactions on Vehicular Technology,~Vol.~XX, No.~XX, XXX~2021}
{}

\maketitle

\begin{abstract}
Controller area networks (CANs) are a broadcast protocol for real-time communication of critical vehicle subsystems. 
Original equipment manufacturers of passenger vehicles hold secret their mappings of CAN data to vehicle signals, and these definitions vary according to make, model, and year.  
Without these mappings, the wealth of real-time vehicle information hidden in the CAN packets is uninterpretable, severely impeding vehicle-related research, including CAN cybersecurity and privacy studies, aftermarket tuning, efficiency and performance monitoring, and fault diagnosis to name a few. 

Guided by the four-part CAN signal definition, we present CAN-D (CAN-Decoder), a modular, four-step pipeline for identifying each signal's boundaries (start bit and length), endianness (byte ordering),  signedness (bit-to-integer encoding), and by leveraging diagnostic standards, augmenting a subset of the extracted signals with meaningful, physical interpretation. 
En route to CAN-D, we provide a comprehensive review of the CAN signal reverse engineering research.
All previous methods ignore endianness and signedness, rendering them incapable of decoding many standard CAN signal definitions. 
Incorporating endianness grows the search space from 128 to 4.72E21 signal tokenizations and introduces  a  web  of  changing dependencies. 
In response, we formulate, formally analyze, and provide an efficient solution to an optimization problem, allowing identification of the optimal set of signal boundaries and byte orderings. 
In addition, we provide two novel, state-of-the-art signal boundary classifiers---both of which are superior to previous approaches in precision and recall in three different test scenarios---and the first signedness classification algorithm, which exhibits a $>$ 97\% F-score. 
Overall, CAN-D is the only solution with the potential to extract any CAN signal that is also the state of the art. 
In evaluation on 10 vehicles of different makes, CAN-D's average $\ell^1$ error is five times better (81\% less) than all previous methods and exhibits lower average error, even when considering only signals that meet prior methods' assumptions. 
Finally, CAN-D is implemented in lightweight hardware, allowing for an on-board diagnostic (OBD-II) plugin for real-time in-vehicle CAN decoding. 
\end{abstract}


\begin{IEEEkeywords}
Controller Area Network (CAN); 
Reverse Engineering; 
Machine Learning; 
Security; 
Privacy; 
Technology; 
\end{IEEEkeywords}



\begin{figure}[!b]
   \centering
    \includegraphics[width=.5\textwidth]{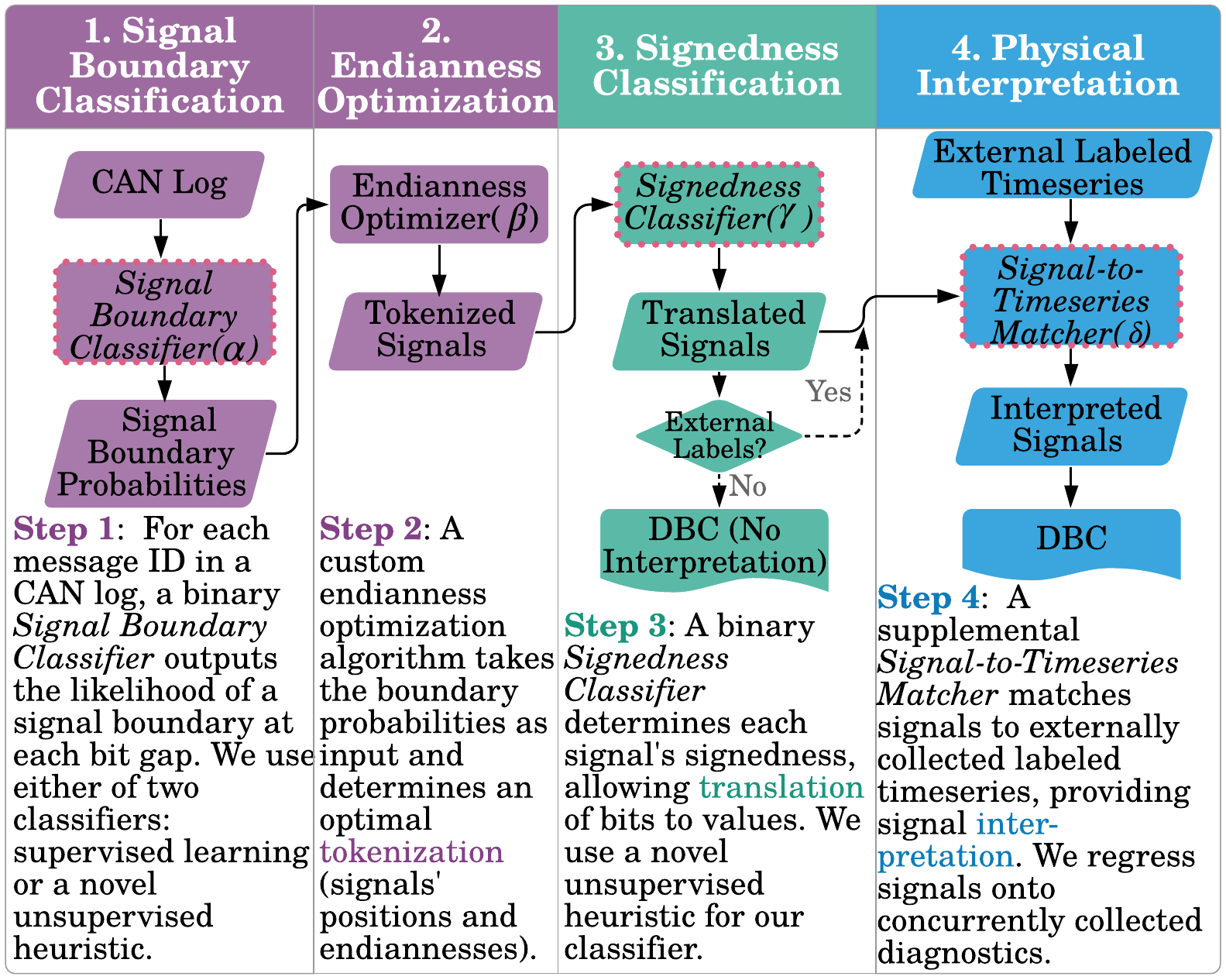}
    \caption{CAN-D Pipeline: A four-step modular pipeline that takes a CAN log (capture of CAN data) as input and outputs a DBC with signal definitions, thus providing vehicle-agnostic CAN signal reverse engineering. Italicized processes outlined in dotted red lines indicate modular pieces that can be any algorithm that satisfies the input/output requirements. Descriptions of our choices for these pieces are provided. Greek letters $\alpha$--$\delta$ denote tuning parameters (possibly) needed for Steps $1$--$4$, respectively.}
    \label{fig:pipeline}
\vspace{-.6cm}
\end{figure}

\section{Introduction \& Background}
\label{sec:intro}
\vspace{-.2cm}
Modern automobiles rely on communication of several electronic control units (ECUs) (internal computers) over a few controller area networks (CANs) and adhere to a fixed CAN protocol. 
Sensors readings, such as accelerator pedal angle, brakes, fuel injection timing, wheel speeds---as well as less important readings, such as radio settings---are all communicated as signals encoded in the CAN messages.  
For passenger vehicles, the encodings of these signals into CAN messages are proprietary: one can monitor (and send) CAN messages but generally cannot understand their meaning. 
Furthermore, these encodings vary by make, model, year, and even trim, and in practice, reverse engineering of signals is currently a tedious, per-vehicle effort. 
Because CAN data is sent at a rapid rate and carries a wide variety of real-time vehicle information, a vehicle-agnostic solution for decoding CAN signals promises a vast resource of streaming, up-to-date information for analytics and technology development on any vehicle.\looseness=-1

Each CAN message has up to 64 bits of data containing (usually) multiple signals (Fig. \ref{fig:dbc_editors}).  
Automotive CAN signals are characterized by four defining properties (discussed in detail in Section \ref{sec:can-basics}): (1) signal boundaries (start/end bit), (2) endianness (byte order), (3) signedness (bit-to-integer encoding), and (4) physical interpretation. The signal definitions for each message (a message definition) are defined in a vehicle's CAN database file (the industry standard is Vector's \texttt{.dbc} or ``DBC'' file format). We use this industry-standard, four-part signal definition to frame our understanding of previous works and guide our approach. 
\textit{The goal is a vehicle-agnostic CAN decoder that can discover these four defining properties for each signal from CAN data from any vehicle---that is, to reverse engineer the signal definitions in the vehicle's DBC. }\looseness=-1

Recently, the research community has focused on reverse engineering signals from automotive CAN data. This research is summarized in Section \ref{sec:related-works}, and Table \ref{tab:related-works} catalogs each work's efforts to identify the four defining signal characteristics.
Notably, all current approaches focus only on identifying signal boundaries (1)
and/or matching signals to observable sensor data (4), while ignoring endianness (2) and signedness (3), meaning they are unable to decode many standard CAN signals.

All previous works have developed and tested algorithms on limited CAN data, often from a single make. 
Targeting a vehicle-agnostic solution, we compile a much more varied collection of labeled CAN data from 10 different makes (see Section \ref{sec:dataset}). 
Equipped with this robust, labeled dataset for development and testing, we pursue the first comprehensive and most accurate signal reverse engineering pipeline (see Fig. \ref{fig:pipeline}). 
Before describing our contributions, we introduce necessary background information.

    \begin{figure}[hb]
    \includegraphics[width=.48\textwidth]{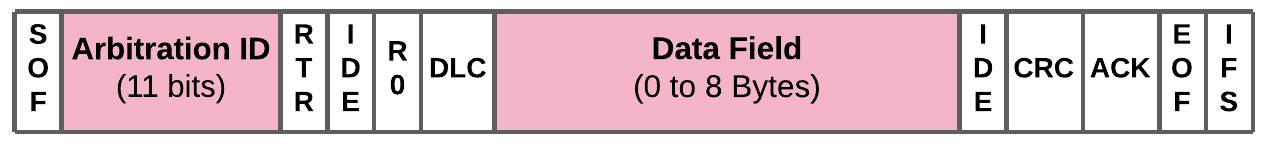}
    \caption{
       CAN 2.0 frame depicted: Arbitration ID indexes the frame; Data Field  carries message content up to 64 bits.
        }
    \label{fig:CANframe}
    \vspace{-.76cm}
    \end{figure}  

\subsection{CAN Fundamentals \& Notation}
\label{sec:can-basics}
\begin{figure}[t]
\vspace{-.3cm}
    \centering
      \includegraphics[width=.48\textwidth]{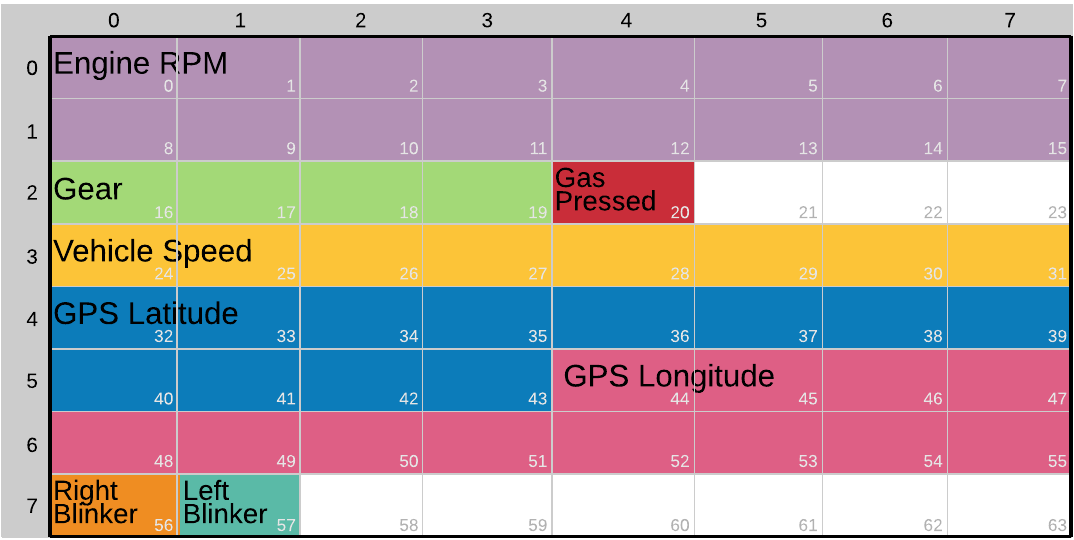}
    \caption{A \textit{signal layout plot} represents a CAN message tokenization, depicting an ID's 64-bit data field as an $8\times8$ array containing CAN signal(s). Each signal's constituent bits are shown in a unique color, and unused bits are shown in white. This representation is similar to visualizations that would be provided by DBC editing software.}
    \label{fig:dbc_editors}
\end{figure}

CAN 2.0 defines the physical and data link layers (Open Systems Interconnection [OSI] layers 1 and 2) of a broadcast protocol \cite{bosch1991can}. 
In particular, it specifies the standardized  CAN frame (or packet) format represented in Fig. \ref{fig:CANframe}.  
For semantic understanding of a CAN frame, only two components of the frame are necessary:
\begin{itemize}[leftmargin = *,topsep=3pt,itemsep=2pt,parsep=2pt]
\item \textit{Arbitration ID}, an 11-bit header  used to identify the frame and  
for arbitration (determining frame priority when multiple ECUs concurrently transmit)
\item \textit{Data Field} or \textit{Message}, up to 64 bits of content
\end{itemize}
Each ID's data field comprises signals of varying lengths and encoding schemes packed into the 64 bits (Fig. \ref{fig:dbc_editors}). 
A \texttt{.dbc} file (standard file type) provides the definitions of signals in the data field for each ID, thus defining each CAN Message.

CAN frames with the same ID are usually sent with a fixed frequency to communicate updated signal values, although some are triggered by an event (e.g., ID \texttt{0x3A2} occurs every 0.1 s, ID \texttt{0x45D} occurs every 0.25 s). 
We partition CAN logs into \textit{ID traces},  the time series of 64-bit messages for each ID. 
An ID trace is denoted  by
$[ B_0(t) \dots, B_{63}(t) ]_t$, a time-varying binary vector of length 64. 
We assume each message is 64 bits by padding with 0 bits if necessary.

\subsubsection{Byte Order (Endianness) and Bit Order} The significance of a signal's bits within a byte (contiguous 8-bit subsequences) decreases from left to right: the first bit transmitted is the most significant bit (MSB), and the last (eighth) bit is the least significant bit (LSB).
This is defined in the CAN Specification \cite{bosch1991can, provencher2012controller} but has been misrepresented \cite{Pese2019librecan} and misunderstood  \cite{nolan2018unsupervised, verma2018actt} by previous signal reverse engineering works. 
The confusion results from use of both big endian and little endian \textit{byte} orderings in CAN messages. Big endian (B.E.) indicates that the significance of bytes decreases from left to right; whereas
little endian (L.E.) reverses the order of the bytes, but maintains the order of the bits in each byte.
For clarity, we list the bit orderings for a 64-bit data field under both endiannesses, and parentheses demarcate the bytes \cite{provencher2012controller}: 
\begin{equation} 
    \begin{aligned} 
    \label{eq:bit-orderings} 
    \hspace{-.25cm} \text{B.E.: }  & (B_{0\phantom{5}}, \dots, B_{7\phantom{5}}), (B_{8\phantom{5}} \dots, B_{15}), \dots , (B_{56}, \dots, B_{63})\\
    \hspace{-.25cm} \text{L.E.: }  & (B_{56}, \dots, B_{63}), (B_{48}, \dots, B_{55}), \dots , (B_{0\phantom{1}}, \dots, B_{7\phantom{1}}) 
    \end{aligned} 
\end{equation} 


\subsubsection{CAN Signals} The specifications for decoding each ID's message into a set of signal values is defined by the original equipment manufacturer (OEM) and held secret. The \texttt{.dbc}  format furnishes the components required to define a CAN signal. Below, the bullet and text colors match those of Fig. \ref{fig:pipeline}.   
\begin{itemize}[leftmargin = *,topsep=3pt,itemsep=1pt,parsep=2pt]
    \item[\textcolor{Plum}{\textbullet}] \textbf{Start bit} and \textbf{length} indicate signal position in the data field.
    \item [\textcolor{Plum}{\textbullet}] \textbf{Byte ordering}: If the signal crosses a byte boundary, \textit{little endian} signals reverse the order of the bytes, whereas \textit{big endian} signals retain byte order (see Eq. \ref{eq:bit-orderings}).
    \item[\textcolor{SeaGreen}{\textbullet}]  \textbf{Signedness}: \textit{Unsigned}, the usual base 2 encoding, vs. \textit{signed}, two's complement encoding \cite{2s-complement}.
    \item[\textcolor{RoyalBlue}{\textbullet}] \textbf{Label} and \textbf{unit} give the physical meaning of the signal and its units (e.g., speed in mph). 
    \item[\textcolor{RoyalBlue}{\textbullet}] \textbf{Scale} and \textbf{offset} provide the linear mapping of the signal's tokenized values to the appropriate units. 
\end{itemize} 

 The above definitional components provide the required learning tasks  for reverse engineering signal definitions. 
\begin{itemize}[leftmargin = *,topsep=3pt,itemsep=1pt,parsep=2pt]
\item \textit{\textcolor{Plum}{\textbf{Tokenization}}}: Learning signals' constituent bits (start bit, length) and byte orders (endianness)---Steps 1 and 2. 
\item \textit{\textcolor{SeaGreen}{\textbf{Translation}}}: Converting a signal's sequence of bits to integers (signedness)--- Step 3. 
\item \textcolor{RoyalBlue}{\textit{\textbf{Interpretation}}}: Linearly mapping (scale, offset) a translated signal to a meaningful measurement (label, unit)---Step 4.
\end{itemize}



\begin{example}
\label{ex:big-lil-endian} 
Consider in Fig. \ref{fig:dbc_editors} the first two-byte \textcolor{Orchid}{\textbf{lavender}}  signal.
To tokenize the signal, or know its sequence (implying order) of bits, we must know endianness. 
If bytes 1 \& 2 are big endian, we obtain MSB-to-LSB bit indices, $I = (0, \dots, 15)$ whereas if they are little endian, the bytes are swapped, obtaining MSB-to-LSB bit indices $I = (8, \dots, 15, 0, \dots, 7)$, notably with $B_{15}\to B_0$.
Next, the signal's signedness furnishes the translation of that bit sequence to an integer. 
The information needed for interpretation is the label and unit of the signal (in this case, engine rpm) and the linear transformation to convert the translated values (a two-byte signal can take $ 2^{16}-1 = 65,535$ values) to the appropriate physical value (e.g., in the $0$--$10,000$ rpm range). 
\end{example}

Fig. \ref{fig:qual_compare_paylods_signals}
illustrates time series of CAN data that have been decoded using both correct and incorrect signal definitions.
Fig. \ref{fig:qual_compare_paylods_signals}(a) plots \textcolor{green}{\textbf{green}} and \textcolor{cyan}{\textbf{blue}} CAN signals tokenized with correct (middle) vs. incorrect (right) signedness, and
Fig. \ref{fig:qual_compare_paylods_signals}(b) plots CAN signals tokenized with correct (top) vs. incorrect (bottom, and in particular, the \textcolor{MidnightBlue}{\textbf{navy}} signal) endianness. The clear discontinuities in these mistokenized and mistranslated signals demonstrate the importance of knowing the endianness and signedness for extracting meaningful time series. 


\subsubsection{On-Board Diagnostics} 
\label{sec:dids} 
In the United States, all vehicles sold after 1996 include an on-board diagnostic (OBD-II) port, which generally allows for open access to automotive CANs, and emissions-producing vehicles sold after 2007 also include a mandatory, standard interrogation schema for extracting diagnostic data using the J1979 standard \cite{J1979_201702}. 
This OBD service  is an application layer protocol by which one can query diagnostic data from the vehicle  by sending a CAN frame. 
A CAN response is broadcasted 
with the requested vehicular state information. 
There is a standard set of possible queries available via this call--response protocol (e.g, accelerator pedal position, intake air temperature, vehicle speed), along with unit conversions, each corresponding to a unique diagnostic  OBD-II PID (DID). 
Specific examples of how to perform the call and response are available \cite{smith_2016, Pese2019librecan}. 
Previous CAN decoding works have iteratively sent DID requests
and parsed the responses from CAN traffic to capture valuable, real-time, labeled vehicle data without using external sensors \cite{huybrechts2017automatic, verma2018actt, Pese2019librecan}.  
$D(t)$ denotes these time series of diagnostic responses, or \textit{DID traces}. 
Inherent limitations exist: the set of available DIDs varies per make, and electric vehicles need not conform to this standard \cite{verma2018actt, Pese2019librecan}.

    \vspace{-.2cm}
\subsection{Problem, Assumptions, and Challenges}
\label{sec:problem} 
\subsubsection{Problem}
The goal is to  to recreate the \texttt{.dbc} file's signal definitions (i.e., discover the four properties for each signal) from a sufficient capture of a vehicle's CAN data, for any vehicle. 

\subsubsection{Assumptions} 
We make five fundamental assumptions: 
\textbf{(A0)} Observed constant bits are unused.\\ 
\textbf{(A1)} Both  big and little endian byte orders are possible.
\textbf{(A1.a)} Both endiannesses can occur in a single ID.  
We have not observed this, but it is permitted by protocol and DBC syntax. DBC editor GUIs allow per-signal endianness specification with a checkbox or pull-down menu, implying that both byte orderings can co-occur in a message. \\
\textbf{(A1.b)} A single byte cannot have  bits used in a little endian signal while also containing bits used in a big endian signal; otherwise, the byte orders indicated by the  signals are contradictory.\\
\textbf{(A2)} Signed signals are possible and are encoded using a two's complement encoding. 

\subsubsection{Challenges} 
In practice, it is difficult to exercise the MSBs of a signal, resulting in errors in determining signal boundaries (a Step 1 challenge). 
For example, consider the two-byte (16-bit) engine rpm signal in Example \ref{ex:big-lil-endian} with translated values of $0$--$10,000$.
As 5,000 rpm is rarely reached, the MSB of this signal is likely to be observed as a constant 0 bit, causing the signal start bit to be mislabeled. 
Though this challenge is easily surmountable for engine rpm (e.g., rev engine in neutral during collection), it is far more difficult to solve this for latent sensors, (e.g, engine temperature). 

Secondly, because continuous signals are sampled periodically, those with high-resolution signals (e.g., a two-byte signal has $2^16 > 65,000$ values) have LSBs flipping seemingly randomly (a Step 1 challenge). 
Our results indicate that the TANG algorithm \cite{nolan2018unsupervised} suffers from the overly strict assumption that flip frequencies decrease monotonically with bit significance. 

Thirdly, considering both big and little endianness greatly increases the complexity of the problem: bits on the byte boundaries have unknown neighbors (albeit in a fixed set of possibilities). For example, simply comparing the bit flip probabilities of neighboring bits now requires custom rules that incorporate all possible neighbors according to  \textbf{(A1)} and \textbf{(A1.a)} given the constraints imposed by \textbf{(A1.b)} (a Step 2 challenge). 

Fourthly, considering both signed and unsigned encodings adds another hurdle. In particular, the order of bit representations mod $2^n$ is the same for both signed and unsigned encodings: half the bit strings represent different integers (a Step 3 challenge). 

Finally, many CAN signals communicate sensor values that are hard to measure with external sensors. Therefore, identifying the physical meaning, the unit and the linear mapping (scale and offset) can be difficult (a Step 4 challenge).

    \vspace{-.2cm}
\subsection{Contributions} 
We make six contributions to the area of automotive CAN signal reverse engineering. 

\textbf{C1. Comprehensive signal reverse engineering pipeline} Our primary contribution is a modular, four-part pipeline, depicted in Fig. \ref{fig:pipeline},  for learning all four components of a CAN signal definition. 
The pipeline is modular in that 
Step 1 can accommodate any signal boundary classification method,
Step 3 can accommodate any signedness classification algorithm,
and Step 4 can accommodate any signal-to-time-series matching algorithm for physical interpretation. 
Instantiating our pipeline with our signal boundary classification heuristic and (separately) our trained machine learning (ML) classifier for Step 1 and the diagnostic sensor matching of Verma et al. \cite{verma2018actt} for Step 4, we present a quantitative comparative evaluation of our signal reverse engineering pipeline vs. previous methods. 
We demonstrate that CAN-D exhibits less than a fifth of the average error exhibited by all previous methods (Section \ref{sec:eval-l1-error} and Table \ref{tb:compare}, bottom), and we qualitatively illustrate the pitfalls and limitations of previous methods (Section \ref{sec:qual_results} and Fig. \ref{fig:qual_compare_paylods_signals}) that our four-step pipeline circumvents. 
\textit{Overall, CAN-D is the first CAN signal reverse engineering effort that can accommodate all signals as defined in automotive DBC files, and it is far more accurate than any previous effort. Furthermore, it provides a framework for future research developments to improve and plug in advancements to each step.}

\textbf{C2. Introduction of two state-of-the-art signal boundary classification algorithms and comparative study of previous algorithms} 
We develop two signal boundary classifiers, a supervised machine learning model, and an unsupervised heuristic (Section \ref{sec:cutprob}). We implement the previous state-of-the-art classification methods 
and provide the first quantitative comparison of all methods (Section \ref{sec:eval-boundary-quant} and Table \ref{tb:compare}, top) on a more comprehensive and robust dataset than any previous work. 
We demonstrate that our algorithms are significantly more accurate than previous methods, superior in both recall and precision in three testing scenarios.

\textbf{C3. Endianness optimization formulation and solution} 
All previous works are \textit{based on an assumption of big endian byte ordering}
(to perform tokenization and/or signal-to-time series matching), and there is no simple remediation for adapting the previous algorithms to perform correctly in the presence of both big and little (reverse byte order) endian signals. 
The second step of our pipeline presents a novel procedure that was crafted to use the predictions from any signal boundary classification algorithm from Step 1 as input and determine the optimal set of endiannesses and signal boundaries from all possible tokenizations (Section \ref{sec:endianess_opt}).
We formulate an objective function to be optimized and provide a formal mathematical proof for reducing the search space to a very tractable grid search algorithm for optimization.
Overall, this insight allows all signal boundary classification algorithms to be leveraged for extracting both little and big endian signals---which has thus far been ignored and/or unachievable. 

\textbf{C4. Signedness classification} 
We provide the first algorithm for determining signal signedness (bit-to-integer encoding) (Section \ref{sec:signed_class}), allowing translation of signals to time series. 
Testing shows this simple heuristic achieves an F-score of $>97\%$.

\textbf{C5. Prototype OBD-II plugin for in-situ or offline use}
The pipeline can be run offline for postdrive analysis  or during driving (to feed online analytics such as a CAN Intrusion Detection System (IDS) with translated CAN data). 
We discuss our design and implementation of a lightweight OBD-II port plugin device (Section \ref{sec:hw} and  Fig. \ref{fig:hw-pic}) for use in any vehicle where a CAN is accessible via the OBD-II port (this is true for most vehicles). 
In a signal learning phase, the device automatically logs CAN data while periodically querying supported DIDs, and then it runs the algorithmic pipeline to learn signal definitions and write a DBC file. 
This allows the real-time decoding of CAN signals on future drives, such as feeding a novel analytic technology leveraging the vehicle's signals online or offline uses, such as  analyzing CAN captures in postcollection analysis. 
This prototype bridges the gap between the algorithmic research in the literature and actual online use with any vehicle. 

\textbf{C6. Survey} We provide the first comprehensive survey of works on reverse engineering CAN signals (Section \ref{sec:related-works} and Table \ref{tab:related-works}), outlining the progression of the field and documenting the benefits and limitations of each contribution.

    \vspace{-.4cm}
\subsection{Impact} 
\label{sec:impact} 

Unveiling CAN signals will provide real-time measurements of vehicle subsystems, offer a rich stream of data that promises to fuel many vehicle technologies, and put development and analytics in the hands of consumers (in addition to OEMs).

Multiple research works---through direct and even remote access to CANs---have managed to manipulate a few manually reverse engineered signals, resulting in life-threatening effects\textemdash most notably, the remote Jeep hack by Miller and Valasek \cite{checkoway2011comprehensive, koscher_experimental_2010, miller_valasek_2014, miller2015remote}. 
These works demonstrate that CAN reverse engineering is possible on a per-vehicle basis with ample effort and expertise and that it will not inhibit the determined adversary. 
However, the obscurity of CAN data does hinder vulnerability analysis research necessary for hardening vehicle systems, 
and automated CAN reverse engineering will greatly expedite vehicle vulnerability research. 

In parallel, CAN defensive security research is growing quickly; we found 15 surveys of the area published since 2017 (e.g.,  \cite{lokman2019intrusion, wu2019survey}) and over 60 works on CAN intrusion detection published between 2016 and 2019.  
Yet these works are impeded by obfuscated CAN data, forced to use either side-channel methods that ignore message contents \cite{moore2017modeling, lee2017otids, choi2018identifying} or black-box methods ignorant of message meanings \cite{Tyree_Bridges_Combs_Moore_2018, Pawelec_Bridges_Combs_2019, taylor2016anomaly},  arduously reverse engineer a few signals for a specific vehicle \cite{narayanan_secure_2016}, or rely on an OEM for signal definitions \cite{hanselmann2019canet}, which keeps CAN security in the OEM's hands and develops per-make, not vehicle-agnostic, capabilities. 
A vehicle-agnostic CAN signal reverse engineering tool promises to remove these limitations and provide rich, online, time series data for advancements in detection and other security technologies. 
Furthermore, this CAN signal decoding will promote universally applicable technologies to address cars currently on the road, as well as remove reliance on the vehicle OEMs for CAN security. 

Another emerging subfield of research is driver fingerprinting \cite{enev2016automobile, wakita2006driver}, developing methods to identify drivers based on their driving characteristics such as braking, accelerating, and steering. 
Access to the decoded CAN data will allow these works to be ported to plugin technologies for nearly any vehicle, impacting, at a minimum, driver privacy and insurance strategies, and potentially forensic (e.g., criminal) investigations and vehicle security to name but a  few benefits.

In addition,  access to CAN signals will potentially assist development of aftermarket tuning tools for enhanced efficiency and performance, fuel efficiency monitoring and guidance, fleet management, vehicle fault diagnosis,  forensics technologies, and aftermarket vehicle-to-vehicle capabilities. 
As a final  example, we note that aftermarket technologies to provide autonomous driving capabilities to current vehicles are appearing. In particular, Open Pilot (\url{https://comma.ai/}) provides latitudinal and longitudinal control for many vehicles on the road using a few presumably manually reverse engineered CAN signals. 
Automated, accurate, and universally applicable CAN de-obfuscation will promote and expedite development of such vehicle technologies, especially aftermarket solutions for many vehicles currently in use.

\section{CAN Signal Reverse Engineering Survey}
\label{sec:related-works} 
This section provides the first comprehensive survey of  methods for decoding automotive CAN data into constituent signals. 
We seek to show the progression of the literature, and we provide more detailed descriptions of the methods that we evaluate in Section \ref{sec:eval} with authors/methods in bold. 
Table \ref{tab:related-works} provides a quick reference for the signal reverse engineering contributions of each work. 


Early work by Jaynes et al. \cite{jaynes_automating_ecu} explored supervised learning to identify CAN messages that control body-related events, but the approach did not account for the fact that data fields comprise multiple disparate signals. Thus, this method simply labels entire messages with a general physical meaning. 

Markowitz and Wool \cite{markovitz2017field} focuses on CAN anomaly/intrusion detection but pursues signal extraction as a preprocessing step. 
They were the first to introduce the basic assumption that each arbitration ID's data field is ``a concatenation of positional [signals].''
Implicitly, Markowitz and Wool's algorithm  assumes only big endian and unsigned signals; thus, their algorithm need only identify the start bit and length of a signal. 
The algorithm considers all 2,080 possible  signals (indexed by start bit and length) in an ID's 64-bit data field, and based on the cardinality of each candidate signal's range, the count of observed distinct values. It then categorizes the signal as constant, categorical (taking on only a few values), or continuous (values of a discretely sampled continuous variable) based on the range and assigns a score. 
Finally, the method identifies a nonoverlapping partition of the 64 bits based on category and a optimization of the signals' scores. 

Huybrechts et al. \cite{huybrechts2017automatic} published the first work to leverage DIDs to annotate CAN data and identify signals. 
Their algorithm converts bytes/byte-pairs in CAN messages to integers and identifies those that are similar to the concurrently collected DID responses, but it operates under the self-acknowledged false assumption that CAN signals are limited to only one- or two-byte signals. 
No linear transformation of extracted signals to the DID sensor values is given.

The next three works, TANG by Nolan et al. \cite{nolan2018unsupervised}, 
ACTT by Verma et al. \cite{verma2018actt},  
and READ by Marchetti and Stabili \cite{marchetti2019read} appear to have occurred independently and concurrently, and we present them chronologically by publication date.

\textbf{Nolan et al.} \cite{nolan2018unsupervised} focus solely on extracting continuous signals by considering transition aggregated $n$-grams (\textbf{TANG}). 
Given an ID trace $[B_0(t), \dots, B_{63}(t)]_t$, Nolan et al. define a TANG vector as $(T_0, \dots, T_{63})$ with 
$T_i = \sum_{t} B_i(t_j) \bigoplus B_i(t_{j+1})$, where $\bigoplus$  denotes XOR. 
Note that this is simply a computationally efficient way to obtain the bit flip count; thus, if an $n$-bit signal's subsequent values change by unit increments, then the LSB will exhibit $T_i = 2^n + 1$, and each next significant bit will have TANG values decreasing by a factor of $2$. 
Roughly speaking, the algorithm for identifying continuous signal boundaries computes the TANG vector from an ID trace, identifies the bit with maximal TANG value as a signal's LSB, and walks left (respectively right for reverse bit order), absorbing bits into the signal until the TANG value increases.
Nolan et al. consider both forward and reverse bit orderings to attempt to take little and big endian encodings into account. 
However, since endianness refers to \textit{byte} (not bit) order, this method cannot accommodate true little endian signals, and in fact violates the fixed bit order defined by the standard. 
Overall, this method assumes big endian, unsigned, and continuous signals.\looseness=-1

\begin{table}[t]
\centering
\caption{Automotive CAN signal reverse engineering algorithms for each of the four signal properties. CAN-D is the only comprehensive algorithm, determining all four properties.}
\label{tab:related-works}
\begin{threeparttable}
\setuptable
\vspace{-.4cm}
\begin{tabular}{lccccccc}
		\multicolumn{1}{l}{}
	    & \headrow{Boundary} 
		& \headrow{Endianness} 
		& \headrow{Signedness} 
		& \headrow{Interpretation}
		\\ 
		\toprule
		Jaynes et al. (2016) \cite{jaynes_automating_ecu}                    &\none	&\none	&\none	 &\prt \\
		Markowitz \& Wool (2017) \cite{markovitz2017field}                   &\full	&\none	&\none	 &\none \\ 
		Huybrechts et al. (2017) \cite{huybrechts2017automatic}              &\prt	    &\none	&\none	 &\prt \\
		Nolan et al.'s \textbf{TANG} (2018) \cite{nolan2018unsupervised}     &\full	&\none	&\none	 &\none \\ 
		Marchetti \& Stabili's \textbf{READ} (2018) \cite{marchetti2019read} &\full	&\none	&\none	 &\none \\
		Verma et al.'s \textbf{ACTT} (2018) \cite{verma2018actt}             &\full	&\none	&\none	 &\full \\
		Pes\'{e} et al. \textbf{LibreCAN} (2019) \cite{Pese2019librecan}     &\full	&\none	&\none	 &\full \\
		Young et al. (2020) \cite{young2020towards}                          &\none	&\none	&\none	 &\prt \\
		\cdashlinelr{1-5}
		\textbf{CAN-D}                                                &\full	&\full	&\full	 &\full \\
		
    \bottomrule
	\end{tabular}
\end{threeparttable}
\end{table}

\textbf{Marchetti and Stabili} \cite{marchetti2019read} propose the Reverse Engineering of Automotive Dataframes (\textbf{READ})  algorithm to extract signals
using heuristics based on a 64-length vector that contains each bit's observed flip probability---that is, $[ P(B_i(t_j) \neq B_i(t_{j+1}) ) ]_{i = 0}^{63}$.   
First, signal boundaries are identified using $m_i := \lceil \log_{10}( P(B_i(t_j) \neq B_i(t_{j+1}) )) \rceil$, the ceiling function of the log probabilities. 
READ follows intuitive logic similar to that of TANG; for continuous signals, an LSB flips \textit{much} more often than an adjacent signal's MSB. Thus, READ places signal boundaries between bits $i$ and $i+1$ iff $m_i > m_{i+1}$---or equivalently if the the bit flip probabilities cross a factor of 10 (e.g., from above $.01$ to below).  
Unlike TANG, READ does not claim to assume only continuous signals, and it in fact builds on Markowitz and Wool's signal categorization efforts. It considers three signal categories---counters (increments by 1 with each message), checksums (hashes for checking whether messages are properly transmitted), and a catch-all bin, ``physical" signals---binning the extracted signals with further heuristics relating to bit flips.
Ultimately, READ partitions an ID's 64-bit data frame into signals with categorical labels. 
The algorithm ignores little endian and signed encoding possibilities and cannot be easily amended to accommodate little endian signals. 
Marchetti and Stabili's evaluations with real and synthetic CAN data compared with Markowitz and Wool's method reveal that READ is far more accurate at finding signal boundaries.\looseness=-1 

\textbf{ACTT} by \textbf{Verma et al.}  \cite{verma2018actt} 
takes an approach fundamentally different from all previous works. Instead of partial tokenization and translation---learning to identify signal boundaries under limiting assumptions (e.g., assuming big endian and unsigned encodings) in an unsupervised fashion---ACTT simultaneously tokenizes, translates, and \textit{interprets} CAN signals. 
The method automatically identifies which DIDs (see Section \ref{sec:dids}) respond on the particular vehicle and then collects ambient CAN data during driving while periodically querying DIDs. 
These diagnostic responses provide labeled time series and DID traces alongside the CAN data, setting up a supervised decoding algorithm. 
For a given ID trace, the constant bits are labeled, and all possible signals (start bit, length) from the remaining nonconstant bits are considered. 
For each possible signal and for each DID trace, linear regression is performed, and a score of linear fit is assigned. 
A scheduling algorithm using dynamic programming then identifies a nonoverlapping set of signals that maximize the fitness score. 
The output is twofold: (1) a list of constant signals, and (2) a subset of signals equipped with linear mappings to a known physical unit that matches a DID  (start bit, length, scale, offset, physical unit, sensor label). 
As in all previous works, this method assumes unsigned encodings, and following TANG by Nolan et al., mistakenly considers reverse bit order as little endian (not \textit{byte} order). 
Because this method relies on DID matching to tokenize signals, only a small subset of signals can be extracted, but all extracted signals are interpretable.\looseness=-1

\textbf{Pes\'{e} et al.} \cite{Pese2019librecan} present \textbf{LibreCAN}, a three-phase process.
(Phase 0) LibreCAN makes tweaks to READ's algorithm for identifying signal boundaries and categorizing extracted signals. 
Specifically, whereas READ identifies signal boundaries by finding where adjacent bit flip probabilities decrease across a multiple of 10, LibreCAN identifies whether adjacent bit flip probabilities drop by a factor of $T_{p0,2}$, a tunable input parameter. 
(Phase 1) LibreCAN next leverages ideas similar to those of Verma et. al. \cite{verma2018actt}, using cross-correlation to match signals to sensor readings from both DIDs and external sensors, then using linear regression to learn the scale and offset. 
(Phase 2) LibreCAN incorporates a novel, semiautomated method for identifying  body-related signals (e.g., door locks, windshield wipers) by filtering IDs based on changes in data fields before and after a user actuates the body-related feature. 
Pes\'{e} et al. note that little endian signals exist, but like all previous methods, their algorithm assumes big endian byte order and unsigned encodings and does not have a natural extension to accommodate little endian signals.

Lestyan et al. \cite{lestyan2019extracting} focus on CAN signal extraction in support of a data-driven driver fingerprinting method. 
Using neighboring bits' probabilities, the authors propose a method to identify signal boundaries yet settle on considering only one- or two-byte signals. 
Supervised learning to match a few  signals (e.g., rpm, speed, and braking) is trained on just one car's (manually labeled) signals  and tested on seven other vehicles. 
This work seems unaware of most other works in the area. 

The most recent CAN reverse engineering work, by Young et. al. \cite{young2020towards}, uses an approach similar to LibreCAN (Phase 2) to match vehicular functions (based on a hand-labeled time series) to CAN IDs using a data change identification algorithm. They use a clustering algorithm to group related IDs, labeling the remaining unknown IDs based on those labeled in the matching step. However, similar to Jaynes et al., this work attempts to assign physical meaning to an entire CAN ID rather than tokenize, translate, and then identify (assign meaning to) constituent signals; thus, we do not consider it (nor the work by Jaynes et al.) to be a true signal reverse engineering algorithm. 

All previous works suffer from significant limitation.
Most notably, all assume both big endian byte order and unsigned encodings.
Although some may theoretically correctly identify signed signals' boundaries, this has not been mentioned or tested. 
Worse, there is no natural extension to little endian and/or signed signals. 
To identify signedness, an additional algorithm is needed\textemdash a fairly straightforward binary classification problem that is not difficult once well formed.
On the other hand, including endianness poses a far harder problem for two reasons: 
(1) signal boundary algorithms depend on flip
counts of 
``neighboring'' bits, but bit orderings change with endianness, so neighboring bits cannot be determined; and
(2) without considering both endiannesses, signal boundary identification is 
computationally simple (the same binary classification is independently repeated 64 times per ID), but considering all byte orderings grows the search space combinatorially ($2^{64}$ boundary options $ \times\ 2^8$ byte orders $> \ 4.72\mathrm{E}21$ tokenizations per ID!) with a web of changing dependencies.

\section{Algorithm}
\label{sec:algo}

We present CAN-D, a four-step modular pipeline (depicted in Fig. \ref{fig:pipeline}) that provides the first comprehensive and vehicle-agnostic CAN signal reverse engineering solution. 
We describe the needed inputs and outputs for the modular components---a \textit{signal boundary classifier} (Step 1, Section \ref{sec:cutprob}), a
\textit{signedness classifier} (Step 3, Section \ref{sec:signed_class}), and a \textit{signal to time series matcher} (Step 4, Section \ref{sec:diag-matching})---as well as our novel \textit{endianness optimizer} (Step 2, Section \ref{sec:endianess_opt}), which we consider to be the unique component that provides the glue for the interchangeable components. 

    \subsection{Step 1: Signal Boundary Classification}
\label{sec:cutprob}
Given an ID trace as input, a signal boundary classifier makes
64 binary classification decisions. It predicts whether each of the 64 bits is the LSB of a signal, effectively deciding whether a signal boundary or ``cut'' occurs between each bit and the following bit.  
Almost all previous works have focused on signal boundary classifiers that use hand-crafted heuristics that leverage only one feature, the probability of each bit flipping.
In this section, we pursue the same goal but use a wider set of features. 
In addition to a novel, unsupervised heuristic,  we leverage supervised ML and deliver two superior signal boundary classifiers. 

For the reverse engineering pipeline, outputs of the signal boundary classifier in Step 1 are inputs to the endianness optimizer in Step 2. 
Although we frame signal boundary identification as a set of binary classifications, the input for Step 2 of the CAN-D pipeline is  
the estimated probability---in $\{0,1\}$ for binary heuristics or in $[0,1]$ for ML---of a signal boundary for each bit. 
Algorithms developed in previous works \cite{markovitz2017field,marchetti2019read, nolan2018unsupervised, verma2018actt} and \cite {Pese2019librecan} (Phase 0) could be used as the signal boundary classifier for this step, all of which produce binary label outputs. 
Section \ref{sec:eval} presents results comparing our signal boundary classifiers against the previous state of the art.


\subsubsection{Data and Notational Setup}  
Both unsupervised and supervised predictions are based on statistics that describe how a particular bit and its neighboring bits flip.  
We use a ground-truth DBC (see Section \ref{sec:dataset}) to create a target vector, 
providing a 0/1 label for each bit to indicate whether it is a signal's LSB (boundary).   
To deal with the issue that neighboring bits at byte boundaries are conditioned on endianness, we split little endian signals on byte boundaries for training the supervised models and testing all models. 
In use, the classifier (heuristic or ML) will be applied to ID traces under both byte orderings (see Eq. \ref{eq:bit-orderings}), creating two sets of predictions. 
Both sets of predictions are input to Step 2, which determines the endianness of each byte. 
\looseness=-1

Here we introduce two views of the data used for training and then scoring/tuning the ML in this section. (Both are also used for testing all methods in Section \ref{sec:eval-boundary-quant}.) 
For training, we remove the constant bits (obvious boundaries) forming a ``condensed trace.''
The motivation for 
this is threefold:
(1) Based on assumption \textbf{(A0)} (see Section \ref{sec:problem}), observed constant bits necessarily delimit signals, so a simple rule suffices to identify these obvious signal boundaries. 
The algorithm's goal is to learn signal definitions from a given set of CAN data; thus, unused bits are limitations in the data that is present. Having ample and varied CAN  data promises to enhance the accuracy of the overall CAN-D pipeline when used for training and to provide more thorough examination of the algorithm's abilities when used in testing. 
(2) Our features encode neighboring bits' values and flips, so when nearby bits are constant, features are either trivial or undefined. 
Removing the constant bits prior to feature building yields a better feature set. 
(3) Classes are highly biased toward the negative class; most bits are not an LSB (not on a signal boundary). 
By removing constant bits, we not only get better features, but we also artificially increase the number of nonobvious signal boundaries and decrease class bias, particularly for the nonobvious examples for which a classifier is needed. Note this is the ``c'' set described in Section \ref{sec:eval-boundary-quant}.
Using this condensed trace, we build a feature array
with shape $m$ nonconstant bits by $n_f$ features (features described below for each method).

Second, for tuning the ML classifiers in this section, we only consider their performance on the nonobvious boundaries in the original data\textemdash those boundaries not abutting constant bits in the noncondensed ID traces. 
Note that this set is the ``f$-$'' set described in Section \ref{sec:eval-boundary-quant}.
We tune our supervised model on this set because we ultimately wish to apply the model to full 64-bit traces and optimize performance for this situation. 

\begin{wraptable}[9]{r}{3cm}
\vspace{-.2cm}
 \fbox{%
   \begin{minipage}{\dimexpr2.9cm-2\fboxsep}
    \small
    \caption{Local bit-flip features; $F_i$ denotes a flip of bit $i$.}
    \label{tb:features}
    \begin{tabular}{l}      
            \toprule
            $P(F_i)$\\
            $P(F_i \mid F_{i+1})$\\
            $P(F_{i+1} \mid F_i)$ \\
            $P(\neg F_i \mid \neg F_{i+1})$\\
            $P(\neg F_{i+1} \mid \neg F_i)$ \\
        \end{tabular}
   \end{minipage}}
\end{wraptable} 
\subsubsection{Supervised Classification}
\hspace{5cm} To describe features conceptually, we use $i\pm1$ to denote bit $i$'s neighbors, notationally neglecting the varying neighbors based on endianness (see Eq. \ref{eq:bit-orderings}) when it presents unnecessary complications. 
For each bit $i$, 
we generate a set of 15 features. 
The first five features are  ``local'' to bit $i$ and its relationship to bit $i+1$, and are denoted by $v^{id}_i \in \mathbb{R}^{5}$.
These features (listed in in Table \ref{tb:features}) are 
estimated probabilities of a ``bit flip'' based on observations in data over time. The flip of bit $i$---alternating value in subsequent messages $B_i(t_j) \neq B_i(t_{j+1})$---is denoted by $F_i$.

The guiding principle is that a signal's LSB generally alternates value much more often than an adjacent signal's MSB; thus, the bit flip features should provide good indicators for boundaries. 
Specifically, the first feature should identify LSBs ($P(F_i) \approx 1$) and MSBs ($P(F_i) \approx 0$). 
This is essentially the feature on which previous works \cite{nolan2018unsupervised, marchetti2019read, Pese2019librecan} base their heuristic. 
The next four conditional bit flip features are expected to differ significantly for adjacent bits contained in the same signal versus those that are part of separate signals because the former are likely dependent while the latter are likely independent. 

Next, we look to the neighboring bit on the right, bit $i+1$, and add the five local features for this bit  $v^{id}_{i+1}$ to our feature set for bit $i$. 
Finally, we add five difference features $\delta(v^{id}_{i+1}, v^{id}_{i})$, yielding a total 15-length feature vector for bit $i$. 

Initially, we experimented with adding a wider variety of features based on bit values, two-bit distributions, and entropy, as well as more left/right neighboring features. However, we found that these features did not improve classification performance and in fact resulted in overfitting.

\begin{table}[t]
    \small
    \centering
    \begin{threeparttable}[]
        \caption{Aggregated classification metrics of signal boundary classifiers scoring only nonobvious boundary decisions (f$-$ set). The best supervised method and unsupervised heuristic (bolded) are used in the CAN-D pipeline. \textbf{Top:} Supervised classifiers are run with default scikit-learn parameters using LOOCV by CAN log. Top-performing models (italicized) are tuned with optimal parameters chosen using a grid search. 
        \textbf{Bottom:} The unsupervised heuristic (see Alg. \ref{alg:boundary-heuristic}) performs only slightly worse than the best supervised method.}%
        \label{tb:boundary_classifier_results}
        \begin{tabular}{lccc}
        \toprule
        {Classifier} &  {F-Score\phantom{m}} &  {Precision\phantom{m}} &  {Recall\phantom{m}} \\
        \midrule
        AdaBoost            &     87.6 &       82.6 &    93.3 \\
        Decision Tree       &     78.5 &       67.8 &    93.3 \\
        \textit{KNN}                 &     88.1 &       81.3 &    96.2 \\
        \textit{KNN (Tuned)}           &     89.7 &       82.6 &    98.1 \\
        Logistic Regression &     86.9 &       82.1 &    92.3 \\
        \textit{MLP\tnote{1}}                 &     88.4 &       82.5 &    95.2 \\
        Naive Bayes          &     71.6 &       57.6 &    94.7 \\
        \textit{Random Forest}       &    90.2 &      85.4 &    95.7 \\
        \textit{\textbf{Random Forest (Tuned)\tnote{2}}}       &    \textbf{ 91.2} &      \textbf{ 87.6} &    \textbf{95.2}\\
        SVC Linear          &     85.5 &       78.6 &    93.8 \\
        \textit{SVC Poly}            &     88.7 &       85.3 &    92.3 \\
        \textit{SVC Poly (Tuned)}      &     88.9 &       85.4 &    92.8 \\
       \textit{SVC RBF\tnote{1}}             &     89.0 &       84.8 &    93.8 \\
        SVC Sigmoid         &     46.4 &       42.3 &    51.4 \\
        \cdashlinelr{1-4}
        \textbf{Heuristic} & \textbf{89.6}    &   \textbf{89.9}        & \textbf{89.4}\\
        \bottomrule
        \end{tabular}
        \begin{tablenotes}
        \small
        \item[1] Default parameters were optimal
        \item[2]\texttt{max\_features=$\sqrt{n_f}$,  
        min\_samples\_Leaf=3, n\_estimators=200,
        max\_depth=5}
        \end{tablenotes}
            
    \end{threeparttable}
\end{table}

We tested the performance of several binary classifiers: 
naive Bayes, logistic regression, support vector classifiers (SVC), decision trees, random forests (RFs), k-nearest neighbors (KNNs), multilayer perceptrons (MLP), and AdaBoost.
After experimenting with different weighting schemes to combat the bias class issue, as well as the fact that we only score the nonobvious boundaries, we settled on a sample weighting scheme of nonobvious-positive:negative:obvious-positive of 8:4:1 (no weighting scheme used for KNN and MLP). 
To test the accuracy of the classifiers, we used leave-one-out cross-validation (LOOCV), holding out one CAN log per fold and aggregating the results, and the f$-$ set, only scoring nonobvious boundaries. 
The results are shown in Table
\ref{tb:boundary_classifier_results}. Focusing only on the best classifiers (those that achieved $\geq 88\%$ F-score using default scikit-learn parameters), we found the optimal parameters for these top-performing  models using a grid search and LOOCV. The best-tuned model is the RF classifier, which yields an overall 88\% precision and 95\% recall for an F-score of 91\%. 
We selected this tuned RF model for our ML classifier.
Finally, as an input to Step 2, we output the classifier's predicted probability of a bit $i$ being a signal's LSB.


    
\subsubsection{Unsupervised Heuristic} 
As an alternative to ML, 
we further explored the feature set to develop a simple heuristic relating to bit flip probabilities.
We found that the conditional bit flip probability $P(F_{i+1}| F_{i})$ and the difference between successive conditional bit flip probabilities $P(F_{i+2} \mid F_{i+1}) -  P(F_{i+1} \mid F_{i})$  are better indicators of a signal ending at bit $i$ than the difference of unconditional bit flip probabilities  $P(F_{i+1}) - P(F_i)$ used by most related works. 



\begin{figure}[t]
    \vspace{-.5cm}  
    \centering
    \includegraphics[width=.48\textwidth]{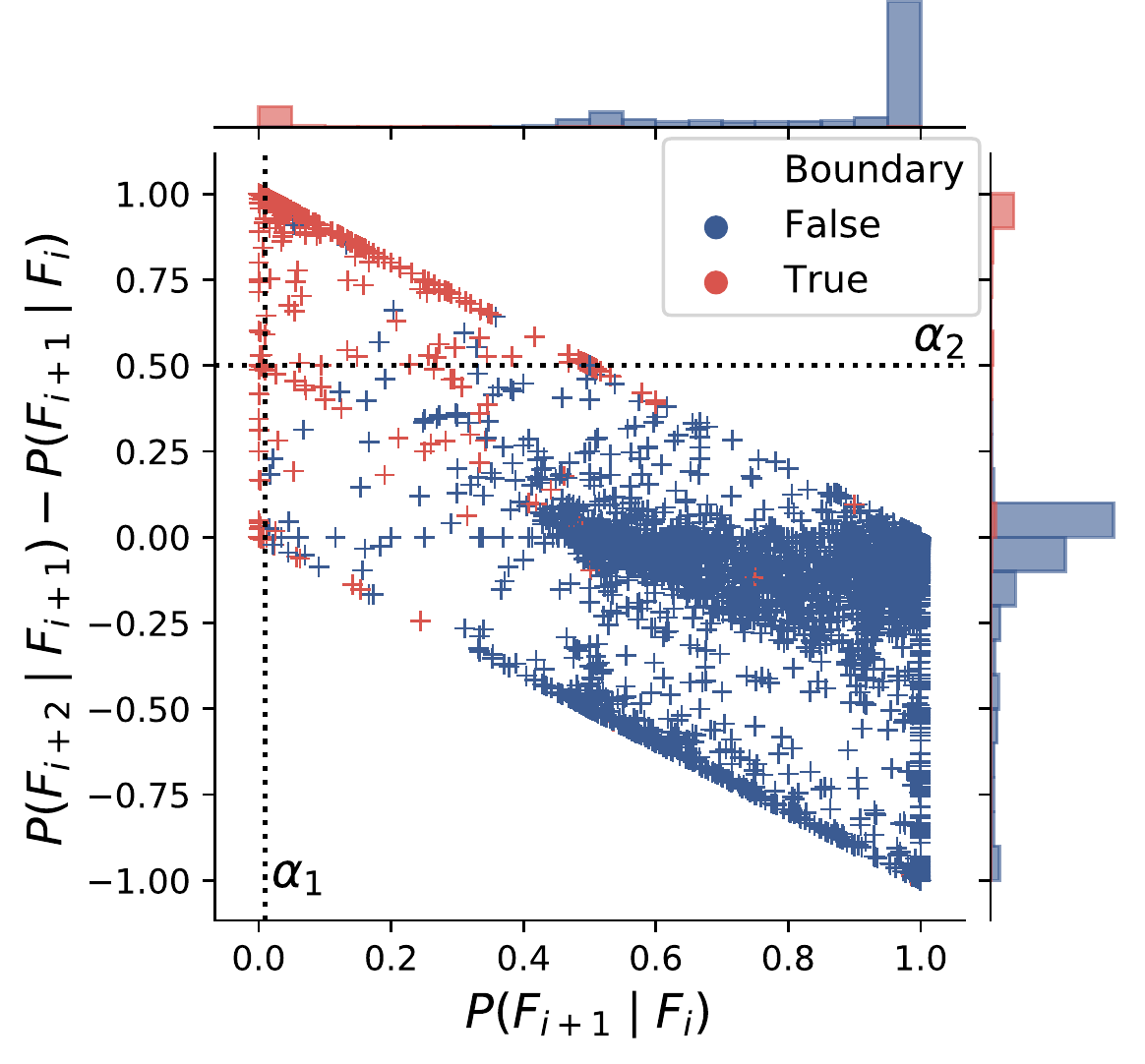}
    \caption{Visualization of heuristic signal boundary classifier (Alg. \ref{alg:boundary-heuristic}) based on conditional bit flip probabilities, with $\alpha_1 = .01$, $\alpha_2 = .5$.}
    \label{fig:boundary_heuristic}
\end{figure}

We developed a heuristic based on these findings, detailed in Alg. \ref{alg:boundary-heuristic} and visualized in Fig. \ref{fig:boundary_heuristic}.
Based on observations of data, we found that setting parameters $\alpha_1 = .01, \alpha_2 = .5$ splits the feature space well and yields a $90\%$ F-score and precision and $89\%$ recall (also on the f$-$ set). 
Note that our heuristic was developed and tuned based on a small preliminary dataset, but we found that it generalized well to all of our data. 

The heuristic's main advantage is that it requires no training while achieving similar accuracy to that of the ML, as shown in Section \ref{sec:eval-boundary-quant}.
Though simple, intuitive, and computationally efficient, one drawback 
is that the outputs are binary labels with no way of determining probabilities properly in $(0,1)$,  
thereby removing some of the flexibility offered by the following step.
\begin{algorithm}[h]
     \caption{Heuristic signal boundary classifier}
     \label{alg:boundary-heuristic}
     \textbf{Inputs} $P(F_{i+1} \mid F_{i})$, $P(F_{i+2} \mid F_{i+1})$, $\alpha_1$ , $\alpha_2$\\
     \eIf{$P(F_{i+1} \mid F_{i}) < \alpha_1$ or $P(F_{i+2} \mid F_{i+1}) -  P(F_{i+1} \mid F_{i}) > \alpha_2$}{
       \Return \texttt{True}
      }{\Return \texttt{False}}
    \end{algorithm}
 \vspace{-.4cm}  
 
    \subsection{Step 2: Endianness Optimization}
\label{sec:endianess_opt}  
Armed with the probability of a boundary or ``cut'' between adjacent bits of a message, we constructed an optimization problem to simultaneously determine the most likely packing of signals into the 64-bit data field and most likely endiannesses of each of the eight bytes. 

\subsubsection{Valid Tokenizations}
$I$ denotes a candidate signal, represented by the list of signal bit indices ordered from MSB to LSB.   
Given a signal $I$, let $LSB(I)$ (or simply $LSB$ if no ambiguity is present)   denote the least significant bit. 
We consider constant bits as 1-bit signals. 
Each ID has eight bytes indexed $j = 0, \dots, 7$ with byte $j$ composed of bits  $8j, \dots, 8(j+1) - 1$  (e.g., byte 2 refers to bits $16, 17, \dots 23.$) Refer to Fig. \ref{fig:endianness_probs} (bottom) payload plot, where each row denotes a byte with labeled bit indices. 
Let $E(j)\in \{B, L\}$ denote that byte $j$ is big, little endian, respectively. 

\begin{defn}[Valid Tokenizations]
\label{def:valid-tokenization}
For a given ID trace, define a \textit{valid tokenization}
$T$ as a tuple of candidate signals $\{I_k\}_k$ and endiannesses of each byte $ \{E(j)\}_{j=0}^7$ such that \\
(1) $\bigcup I_k = \{$0, \dots, 63$\}$ (all 64 bits are used),\\
(2) $I_k \bigcap I_l = \emptyset$ for all $k\neq l$ (signals do not overlap), \\
(3) Assumption (A1.b), one endianness per byte, is satisfied (implicit in the notation $E(j)$). 
\end{defn}

\begin{example}
\label{ex:boundary} 
Consider Fig. \ref{fig:endianness_probs} (bottom), a signal plot layout depicting a valid tokenization with one color per signal (and constant bits in grey).  
The \textcolor{Blue}{\textbf{navy}} signal, a 10-bit little endian signal starting at bit 0, is denoted 
$I = ({14}, {15}, 0, \dots, 7)$. 
Because, $B_{15}\to B_0,$ necessarily $ E(0) = E(1) = L$. 
\end{example}

Example \ref{ex:boundary} shows that if a signal $I$ crosses a byte boundary, the endianness of both bytes  is determined by the order of the indices according to Eq. (\ref{eq:bit-orderings}). 
This leads to the following definition and proposition, which will play an important role in the computational tractability of our optimization problem. 

\begin{defn}[Byte Boundaries]
\label{defn:bytejoins} 
For $j = 0, \dots, 7$ let $v(j) \in \{J_B, J_L, C\}$ denote whether byte boundary $j$ is
\begin{itemize} 
\item a cut $(C)$: bit $[8(j+1)-1]$ ends a signal or is constant,
\item a big endian join ($J_B$): $[8(j+1)-1]\to 8(j+1), $ or
\item a little endian join $(J_L)$: $[8(j+1)-1]\to 8(j-1)$
\end{itemize}
and $V:= \{v \in \{J_B, J_L, C\}^8 \mid v$ is valid byte boundary set$\}.$
\end{defn}

For bits not on a byte boundary ($i\notin S:= \{8j-1\}_{j=0}^7$), there are only two options: cut or join $B_i \to B_{i+1}$. Both are valid possibilities regardless of endianness. 

\begin{prop} 
\label{prop:joins}
A valid tokenization $T$ has $v$ satisfying: 
\begin{enumerate}
\item $v(j) = J_B \implies 
 E(j) = E(j+1) =  B $
\item $v(j) = J_L \implies 
E(j-1) = E(j) =  L $
\item $v(0) \neq J_L$ 
\item $v(7) \neq J_B$
\item $v(j) = J_B \implies v(j+1) \neq J_L, v(j+2) \neq J_L$
\end{enumerate}
\end{prop}
\begin{proof}
(1) and (2) follow directly from Eq. (\ref{eq:bit-orderings}) (endianness definition) and assumption \textbf{A1.b} (one endianness per byte). 

For (3), $v(0) \neq J_L$ else $0\to -8\notin [0,63]$---similarly for (4). 

For (5), if $v(j) = J_B$ and either $v(j+1) = J_L$ or $v(j+2) = J_L$, then (1) and (2) imply $E(j+1)$ is both big and little endian, a violation of assumption \textbf{A1.b}. 
\end{proof} 
\vspace{-.25cm}
\begin{rmk}
\label{rmk:v}
Proposition \ref{prop:joins} can be summarized by defining $V:= \{v\in\{J_B, C\} \times \{J_B, J_L, C\}^6 \times \{J_L, C\}$ with no consecutive subsequences of the form $(J_B, J_L)$ or $(J_B,  * , J_L) \}$.
\end{rmk}

\begin{defn}[$\mathcal{T}$ \& $\mathcal{T}_v$]
Let $\mathcal{T}$ denote the set of valid tokenizations.
For $v\in V$  let  $\mathcal{T}_v \subset \mathcal{T}$ be the tokenizations with byte boundaries defined by $v$.  
\end{defn}

\begin{cor}
\label{cor:counting} 
There are 
$|\mathcal{T}| =  |V| \times |\mathcal{T}_v| = 577\times 2^{64-8} \ \approx 4.16\mathrm{E}19$ valid tokenizations. 
\end{cor}
\begin{proof} 
 $| \{J_B, C\} \times \{J_B, J_L, C\}^6 \times \{J_L, C\}| = 2^2 \times 3^6,$ and removing subsequences of the form $(J_B, J_L)$ or $(J_B,  * , J_L)$ leaves 577.  $|\mathcal{T}_v| = 2^{64-8}$ as the remaining $64-8 $ bit gaps have two valid options: cut or join. 
\end{proof} 
\vspace{-.25cm}


\subsubsection{Optimization Formulation}
The output from Step 1 provisions $f(i | E(j_i))$ = $P ($cut to the right of bit $i$ for endianness $E(j_i))$, with $j_i = \lfloor i/8 \rfloor$ denoting the corresponding byte index for bit $i$. 
We set $f(i, e) = \infty$ if bit $i$ is to the left of a mandatory cut (e.g., the next bit is a constant bit).   
For intuition in the formulation below, consider $f(i| E(j_i) )$, not as the likelihood of a cut, but as penalty for not cutting, and let $\beta$ be a fixed cut penalty parameter. 

The idea for our cost function is to let signals accrue a join penalty---the sum of the probabilities $f(i|E(j_i))$ for each bit that is not cut in order to form the signal. 
Because the candidate signal entails a cut to the right of its LSB, we swap $f(LSB, E(j_i))$ for $\beta$, the cut penalty.
Thus, the $\beta$ controls how liberal to be with cuts. 

The intuition is to find the optimal balance between partitioning the message into too many signals and joining multiple disparate signals by balancing the cut penalty ($\beta$) with the likelihood of a cut (join penalty $f$). 
Setting $\beta = 1$ will lead to cutting only where $f(i|\cdot) = \infty$ (signals demarcated by constant bits), and $\beta = 0$ will lead to a cut at every gap, resulting in 64 1-bit signals. 


\begin{defn}[Costs]
Define the \textit{Signal Cost} as 
\begin{equation*} 
\label{eq:signal-penalty} 
\phi (I, E) :=   
\underbrace{ \sum_{i \in I\backslash\{LSB\}} f(i |E(j_i)) }_{\text{\textit{join penalty}}} \  + \underbrace{\vphantom{\sum_{i \in I \backslash \{LSB\}}} \beta}_{\text{\textit{cut penalty}}}.
\end{equation*} 
Extending to a Tokenization Cost, we have
\begin{align*}
\label{eq:schema-cost} 
\Phi(T):    &= \sum_{I \in T} \phi(I,E) \\
            &= \sum_{\chi_T(i) = 0} f(i| E(j_i)) + \sum_{\chi_T(i) = 1} \beta\\
            &= \sum_{i=0}^{63} (1-\chi_T(i))f(i| E(j_i)) + \chi_T(i) \beta
\end{align*}
with $\chi_T(i) = 1 $ if $i$ is an LSB of a token in $T$, else $0$. 
\end{defn}

The above definition sets up our optimization problem: identify the \textit{optimal tokenization}
\begin{equation}
\label{eq:optimization-problem}
  T_0 := \underset{T \in\mathcal{T}}{\arg\min }\  \Phi(T) \; .
\end{equation}

\begin{figure}[t]
    \centering
    \includegraphics[width=.21\textwidth]{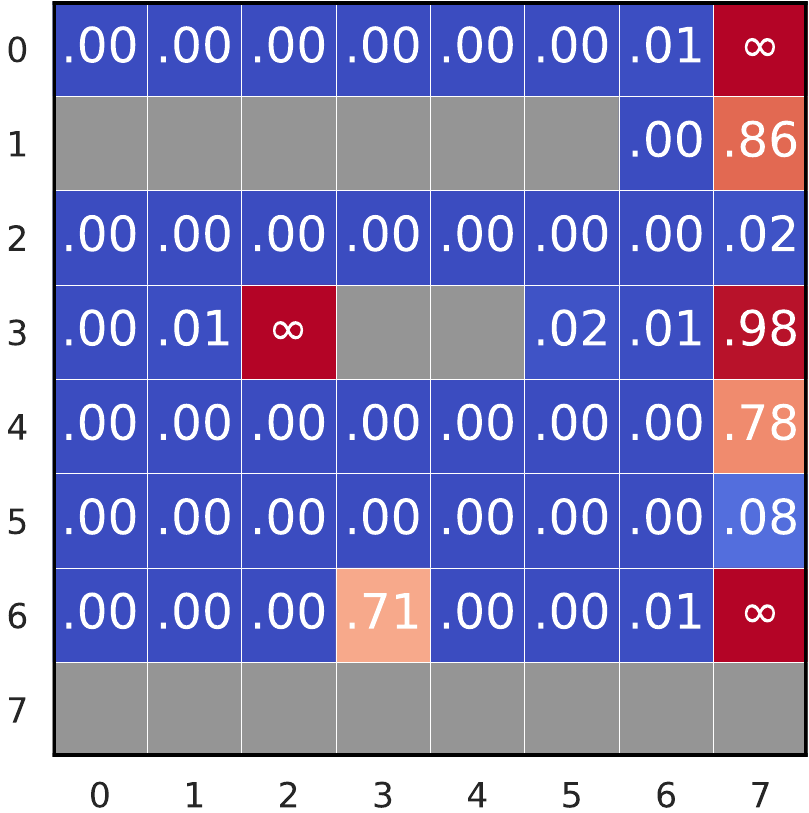}
    \includegraphics[width=.21\textwidth]{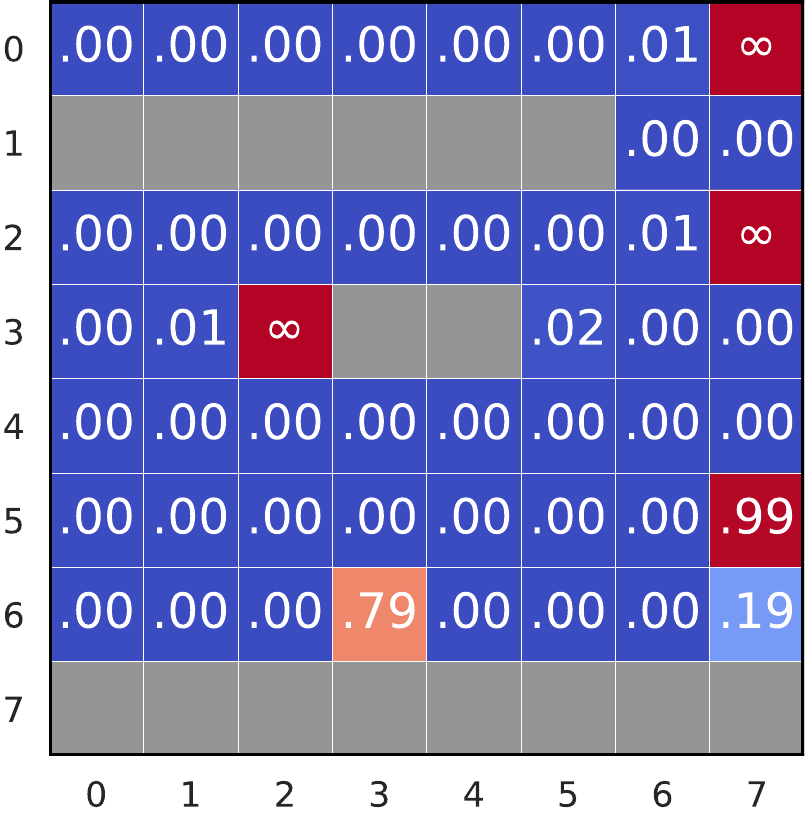}
    \includegraphics[width=.25\textwidth]{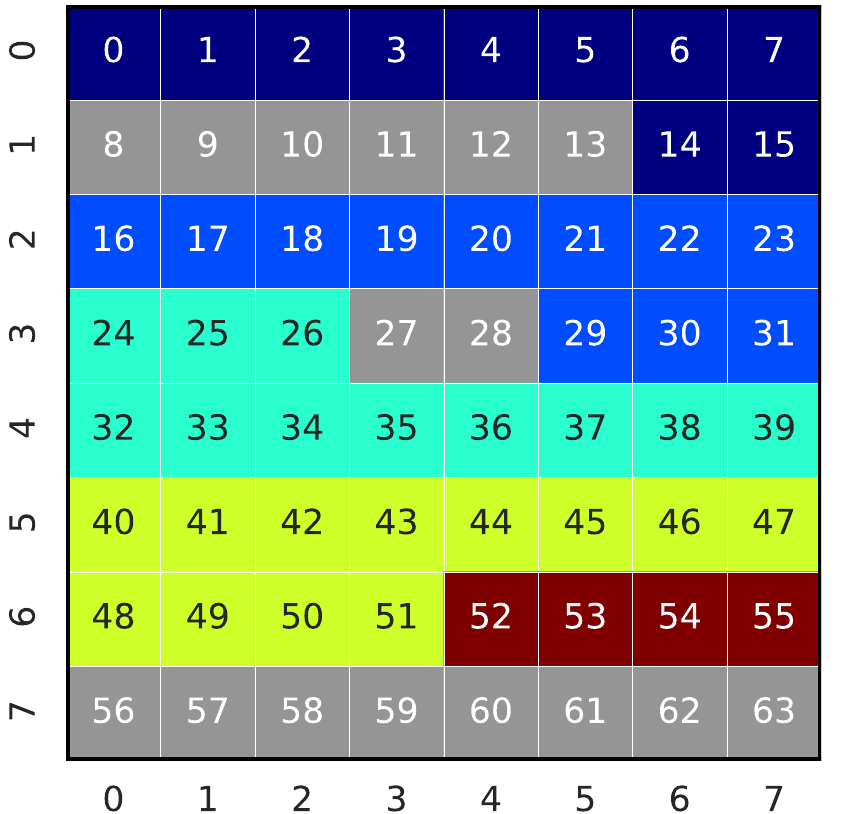}
    \caption{\textbf{Top:} Probabilities of boundaries according to big endian ordering \textbf{(left)}, little endian ordering (\textbf{right)}. \textbf{Bottom:} The resulting optimal tokenization using $\beta=.6$ is three little endian (\textcolor{Blue}{\textbf{navy}}, \textcolor{Cerulean}{\textbf{blue}},  \textcolor{SeaGreen}{\textbf{teal}}), one big endian (\textcolor{SpringGreen}{\textbf{lime}}) and a 4-bit (\textcolor{Maroon}{\textbf{maroon}}) signal.}
    \label{fig:endianness_probs}
\end{figure}

\begin{example}
Refer to Fig. \ref{fig:endianness_probs} and
consider using the cost function to make a local endianness decision for byte 4, comparing the two overlapping 11-bit candidate signals that both contain byte 4 (bits 32 to 39 as numbered in the bottom 
plot): 
a big endian signal $I_0 = (29, \dots, 31, 32,\dots, 39)$ and a little endian signal $I_1 = (32, \dots ,39, 24,\dots, 26)$.  
Using the big $f(\cdot | E = B)$ and little $f(\cdot | E = L)$ endian probabilities depicted in the left and right plots, respectively, penalties for these candidate signals are $\phi_{\beta, f}(I_0, B ) = .02+.01+.98 + \beta = 1.01 + \beta$ and  $\phi_{\beta, f}(I_1, L) =.01 + \beta$. Because $ .01 + \beta < 1.01 + \beta$, $(I_1, L)$ has a lower penalty (regardless of the choice of $\beta$), it is locally optimal. 
In fact, $(I_1, L)$ turns out to be in the globally optimal $T_0$, which is shown in the bottom plot in \textcolor{SeaGreen}{\textbf{teal}}. 
\end{example}

\vspace{-.2cm}


\subsubsection{Finding an Optimum} 
Given a cut penalty $\beta\in [0,1]$ and 
precomputed cut probabilities---$f(i | E(j_i) )$ for all $i \in\{0, \dots, 63\}$ and both endiannesses $E(j_i)$ (see Step 1, Section  \ref{sec:cutprob})---our goal is to identify an optimal tokenization (Eq. \ref{eq:optimization-problem}) from the $4.2\mathrm{E}19$ valid options.

\begin{thm}
\label{thm:optimization} 
Fixing $v\in V$, where $v$ gives cuts/joins at byte boundaries (bits in $S = \{8(j+1)-1\}_{j=0}^7$), the subproblem
$$\underset{T \in\mathcal{T}_v}{\arg\min }\  \Phi_{\beta, f}(T) $$ is realized by $T_{0,v}$, the tokenization: for all $i\in [0,63]\backslash S$, bit $i$ is an LSB (cut to the right of bit $i$) iff $\beta < (f(i|E(j_i))$.
\end{thm} 
\vspace{-.5cm}
\begin{myproof}
Let $T_{0,v}$ be defined as above and $T\in\mathcal{T}_v$. By definition, for $i\notin S,\ T$ will accrue cost $\min(f(i| E(j_i)), \beta)$. 
Because $T, T_{0,v} \in \mathcal{T}_v$ both accrue the same cost for bits $i\in S,$
it follows that
\begin{alignat*}{2} 
\Phi(T)-\Phi(T_{0,v}) &= \sum_{i \notin S} && \big [(1-\chi_T(i))f(i| E(j_i))  \\
& && + \chi_T(i) \beta - \min(f(i| E(j_i)), \beta) \big ] \\ 
&\geq 0 \; . && \phantom{qedspaceqedspaceqedspace} \qedhere
\end{alignat*}
\end{myproof}

This gives an efficient, constant time search algorithm (689 operations) by (1) storing the optimal cut/join choice for each bit $i\in [0,63]\backslash S$ under each endianness ($56\times 2$ operations), then (2) applying Theorem \ref{thm:optimization} to realize both $T_{0,v}$ and cost $\Phi(T_{0,v})$ for each of 577 $v\in V$ and maintaining the minimum. 

In the case that there are multiple optimal tokenizations, we break ties by choosing the one with the maximum number of cuts, followed by the minimum number of little endian signals, which necessarily furnishes a unique optimal solution. This tie-breaking scheme was chosen because we desired a scheme that involved a simple local decision that favored cutting over joining (less information loss) and big endian signals over little (since big endian signals are far more common). Experimentation indicated that prioritizing the former rule over the latter yielded better results. 

After experimenting with adjusting the tuning parameter $\beta$, we found that $\beta \in [.5, .7]$ yielded fairly consistent and correct tokenizations, so for our pipeline we chose $\beta=.6$. 
Note that the heuristic classifiers in Step 1 provide probabilities in $\{0,1\}$, meaning all choices of $\beta$ yield identical results. Furthermore, note that with binary inputs, a tie-break scheme is often necessary, whereas with high-precision probability inputs, multiple optimal tokenizations with the same cost are virtually impossible. 

The outputs of the endianness optimizer described in this step are tokenized signals. Although in theory another endianness optimizer could be developed and exchanged for this component, we consider this custom optimization to be a fixed and noninterchangeable component of the pipeline. 

    \subsection{Step 3: Signedness Classification}
\label{sec:signed_class}
\vspace{-.15cm}
A signedness classifier takes a tokenized signal (start bit, length, endianness) and makes a binary decision on whether each signal of length greater than two is signed (using two's complement encoding) or unsigned. 
To develop our classifier, we followed a similar workflow to Step 1 (Section \ref{sec:cutprob}), experimenting with supervised classifiers and unsupervised heuristics. 
Because each signal is tokenized, and thus the LSBs and MSBs are now known, this problem is significantly simpler, and features can be developed per signal rather than per bit. 
However, after experimenting with several features and supervised classification methods, we found that a simple heuristic based on the the distribution of the two MSBs of the signal yielded better results than the supervised methods. Using this heuristic, described in Alg. \ref{alg:signed-heuristic}, we obtained almost perfect classification ($97.3\%$ F-score), so we chose to use this heuristic in the CAN-D pipeline rather than a learned model. 

The heuristic is based on how the two MSBs will behave if the signal is signed or unsigned. 
Let $B_{i_0}, B_{i_1}$ denote the MSB and next most significant bit of the signal. 
First, consider the probabilities of the center values, $P[(B_{i_0}, B_{i_1}) = (1,0)]$,$P[(B_{i_0}, B_{i_1}) = (0,1)] $. 
If a signal is signed, values close to zero $(B_{i_0}, B_{i_1})$ will be $(0,0)$ (small positives) or $(1,1)$ (small negatives), whereas values near extremes will be $(1,0)$ (near min) or $(0,1)$ (near max). 
A signal with a small probability of these values is therefore likely signed.
Second, consider the probability of a jump  between extreme values, $P[(B_{i_0}(t_j), B_{i_1}(t_j)) = (0,0) \wedge (B_{i_0}(t_{j+1}), B_{i_1}(t_{j+1})) = (1,1)]$.
If a signal is signed, the two MSBs must flip from $(0,0)$ to $(1,1)$ when changing from a small positive  to a small negative value. 
However, if it is unsigned, flipping is unlikely to ever happen because it would entail changing from a very small value to a large one, resulting in a significant discontinuity. 
If this probability is 0, the signal is likely unsigned. We apply these two ideas as described in Alg. \ref{alg:signed-heuristic}: we set $\gamma = .2$ based on observations of data. 

After Step 3, Signedness Classification, each ID's  64-bit message is partitioned into signals, of which we know the start bits, lengths, endianness, and signedness.
Consequently, each signal can now be translated 
into a time series of integers, denoted by $s(t)$. No previous works have attempted signedness classification, so the signedness classifier presented in this section is currently the sole option for this modular component.

\begin{algorithm}[t]
\small
 \caption{\small Heuristic signedness classifier}
  \label{alg:signed-heuristic}
 \textbf{Inputs}: $\{B_{i_0}(t), B_{i_1}(t)\}_t$, $\gamma$\\
 \If{$P[(B_{i_0}, B_{i_1}) = (1,0)] + P[(B_{i_0}, B_{i_1}) = (0,1)] = 0$}{
   \Return \texttt{True}
  }
  \If{$P[(B_{i_0}(t_j), B_{i_1}(t_j)) = (0,0) \wedge$ $(B_{i_0}(t_{j+1}), B_{i_1}(t_{j+1})) = (1,1)] = 0$}{
   \Return \texttt{False}
  }
    \If{$P[(B_{i_0}, B_{i_1}) = (1,0)] + P[(B_{i_0}, B_{i_1}) = (0,1)] < \gamma$}{
   \Return \texttt{True}
  }
  \Return \texttt{False}
\end{algorithm}
    \subsection{Step 4: Physical Interpretation}
\label{sec:diag-matching}

For our signal to time series matcher, we followed ACTT by Verma et al.  \cite{verma2018actt} to match a subset of the translated signals with diagnostic data. 
This augments matched signals with the necessary information to interpret them as actual measurements in the vehicles. 
We do this by comparing each signal time series $s(t)$ to each DID trace $D(t')$ and determining whether they are linearly related. 
Because the DID traces are sampled at a lower rate than normal CAN traffic, we interpolate the signal values over the diagnostic timepoints, obtaining $s(t')$. 
We then regress $D(t')$ onto $s(t')$ and find the best linear fit, furnishing the coefficients $a, b$ so that $\bar{s}(t) := a s(t') + b \approx D(t')$. 
To score the model's fit, we use coefficient of determination $R^2$, which measures the fraction of total variation in time series $D(t')$ that is explained by $\bar{s}(t')$.
Thus, $R^2 = 1$ exhibits a perfect fit, and $R^2 = 0$ exhibits the fit of a horizontal line (assuming $D(t)$ is not the horizontal line). 
For each signal $s$, we find the diagnostic $D$ that yields the highest $R^2$ value. 
If $R^2 > \delta$, where $\delta \in [0,1]$ is a tuning threshold, $s$ is matched to $D$.
A $\delta = 1$ will return only perfectly matched signals, whereas a small $\delta$ will allow for signals with a lower $R^2$ score to be matched. 
For our implementation, we chose $\delta=.5$. We echo Verma et al.~\cite{verma2018actt} that $\delta$ ``can be tightened to isolate near-perfect encodings, or tuned down to discover multiple related signals to each DID.''

It important to note that this matching is based on translated values and thus is independent of how the labeled time series (diagnostics) and signals are formatted or scaled. We reiterate that this means that the matching algorithm is able not only to recover signals that are sent in normal CAN traffic in addition to diagnostics (although often at much higher precision in the normal traffic), but also can recover signals that are not available in the diagnostics at all. For example, a signal relaying a binary indicator of brake pedal depression (not a diagnostic) may match to accelerator pedal position (DID 73) with a negative linear coefficient $a$. 

 
For signals that match a diagnostic, we have interpretation,
having procured the label and units, as well as the scale $a$ and offset $b$. 
In additional to ACTT \cite{verma2018actt}, LibreCAN \cite{Pese2019librecan} (Phase 2) proposes signal to time series matching algorithms that could be used interchangeably (or even combined) for this component. 
Finally, note that translated signals that are not augmented with labels through this physical interpretation step are still highly valuable because there are many applications in which these unlabeled translated time series are far more useful than binary data.

 \section{Dataset}
\label{sec:dataset}
Because our goal is to build a vehicle-agnostic signal-extraction capability, we collected CAN data from 10 different vehicle makes and years ranging from 2010 to 2017 for training and evaluation. 
The details of defined signals for each log are described in Table \ref{tb:log_signal_stats}.
This dataset is far larger and more varied than any previous work. 
Notably, in order to test generalizability of the methods, no duplicate makes were included because different models of the same make (e.g., Toyota Camry and Corolla) have similar characteristics. 


To obtain  data for our signal reverse engineering process (bit position, endianness, and signedness), we used DBCs acquired  from two sources. 
Log \#1 is from a vehicle that uses the J1939 standard \cite{J1979_201702}, a protocol for heavy trucks that provides signal definitions that are publicly available. We were therefore able to obtain absolute ground truth labels for the signals in this log. Because this log contains every type of signal (little endian, big endian, signed, unsigned) and we have absolute confidence in these labels, we consider this log to be the gold standard for testing. 

For logs \#2--10, we made use of CommaAI's OpenDBC project.\footnote{\url{https://github.com/commaai/opendbc}} 
This is an open, crowdsourced set of DBCs that was constructed by individuals using a CommaAI Panda device (an OBD-II plugin) along with the CommaAI Cabana interface to hand-label the data for their vehicle through trial, error, and visual inspection. 
OpenDBC includes DBCs for only a limited number of vehicles, and only a subset of IDs/signals for each vehicle is defined (see difference in IDs and Def. IDs in Table \ref{tb:log_signal_stats}). 
In particular, unobservable signals are often missing because these are created based on visual inspection. 

\begin{table}[t]
    \centering
    \caption{Statistics on 10 CAN logs, each collected from a vehicle of a different make.  For each log, we enumerate the following: nonconstant IDs (\textbf{IDs}), nonconstant IDs defined by CommaAI (\textbf{Def. IDs}), and each of the encodings of defined signals (big/little endian, signed/unsigned) resulting from ground truth labeling process (see Sec. \ref{sec:dataset}). Three logs contain a high percentage of little endian signals, and all but one contain signed signals.}
    \label{tb:log_signal_stats}
    \begin{threeparttable}
    \begin{tabular}{c r rrr rrr rrr r r}
    \toprule
    \textbf{Log} &  \textbf{IDs}\tnote{1} & \multicolumn{3}{c}{\textbf{Def. IDs}} & \multicolumn{3}{c}{\textbf{Unsigned, B.E.}} &  \multicolumn{3}{c}{\textbf{Signed}} &  \textbf{L.E.} &  \textbf{Total} \\
    \midrule
    \#1\tnote{2} &    54 &                 &17& &              &61& &       &3& &    25 &     89 \\
    \#2  &    66 &                          &14& &            & 143 & &      &21& &     0 &    164 \\
    \#3  &    35 &                          &7& &             & 50& &      &18& &     0 &     68 \\
    \#4  &    79 &                         &28& &             &181& &       &0& &     0 &    181 \\
    \#5  &    63 &                         &21& &             &111& &       &5& &     0 &    116 \\
    \#6  &    22 &                         &19& &              &72& &       &2& &    14 &     87 \\
    \#7  &    26 &                          &8& &              &53& &       &3& &     0 &     56 \\
    \#8  &    40 &                          &8& &              &98& &      &10& &     1 &    108 \\
    \#9  &    27 &                         &17& &              &56& &       &7& &    18 &     75 \\
    \#10  &    55 &                         &14& &             &136& &      &21& &     0 &    157 \\
    \bottomrule
    \end{tabular}
    \begin{tablenotes}
    \small
    \item[1] Nonconstant IDs: IDs with more than one nonconstant bit
    \item[2] Vehicle adheres to J1939 Standard protocol \cite{J1939_201308}, and signal definitions are derived from this open standard.
    \end{tablenotes}
    \end{threeparttable}
\end{table}


We collected data from vehicles that had a closely matching CommaAI DBC (same manufacturer, similar model/year/trim). 
Although CommaAI appears to do some measure of quality assurance on this data, because of the crowdsourced nature of the data, we must be mindful of the potential for mislabeled signals.
We performed quality control on this data using the following process. 
First, we partitioned each nonconstant ID trace into sequences of contiguous, nonconstant bits and label each as an unsigned, big endian signal. 
This provides a set of ``baseline'' signal definitions. 
Next, we parsed the data according to the DBC, trimming off any signal MSBs that were constant in our data because of extreme values not being reached, and redefined the signal to have the trimmed start bit and length. 

For IDs defined by CommaAI, we compared the baseline and trimmed CommaAI signal definitions.
If the definitions for the ID agree and pass a visual check,  we automatically added them to our ground truth DBC.
If they disagreed, we displayed the signal tokenization layout and signal time series' plots (see Fig. \ref{fig:qual_compare_paylods_signals}) side by side and identified the discrepancies.
If easily resolvable by experimentation and visual inspection (e.g., adding an obvious signal definition missing from CommaAI DBC, identifying an unsigned signal mislabeled as signed), we added the correct signal definitions to our DBC with minor errors fixed. 
If we were unable to resolve the discrepancy easily, we simply did not include the ID or signal in our ground truth DBC. 
For IDs that did not appear in the CommaAI DBC, we performed a similar process of visual inspection of the baseline plots, adding these definitions to our ground truth if easily resolved and discarding them if not. 

\begin{example}
For an example of this visual inspection process, consider Fig. \ref{fig:qual_compare_paylods_signals}(c), bytes 5 and 6. Suppose the trimmed CommaAI and baseline signal definitions yielded the top payload/signal plots (\textcolor{Maroon}{\textbf{maroon}} signal) and bottom payload/signal plots (\textcolor{Maroon}{\textbf{maroon}} \& \textcolor{Orange}{\textbf{orange}} signals), respectively. By inspection, we can clearly see that trimmed CommaAI definition (top) is correct (a two-byte little endian signal) because it results in a clear continuous  (\textcolor{Maroon}{\textbf{maroon}}) signal. Note that the baseline definition visualization is characteristic of a truncated signal's LSB (intermittently noisy/constant \textcolor{Orange}{\textbf{orange}}) and a signed signal (\textcolor{Maroon}{\textbf{maroon}} discontinuous near extreme values).
Thus, we would add the correct trimmed CommaAI definition. 
\end{example}

This process of visual inspection method is tedious and time consuming, but it is quite effective as  both legitimate signals and misclassified encodings (e.g., signed signals translated incorrectly as unsigned) 
are usually recognizable from time series plots. 
In fact, our visual inspection process is quite similar to the CommaCabana interface used to create CommaAI DBCs. 

\section{Evaluation}
\label{sec:eval}
Using the dataset described in Section \ref{sec:dataset}, we compared our algorithms, both with the heuristic\footnote{$\alpha_1 = .01, \alpha_2 = .5, \beta = .6$ (though irrelevant with binary Step 2 inputs),$ \gamma = .2, \delta = .4$.} and ML\footnote{Step 1 with tuned RF model found in Sec. \ref{sec:cutprob}, $\beta = .6, \gamma = .2, \delta = .4$.} for Step 1, against the following predecessors:  
TANG \cite{nolan2018unsupervised}\footnotemark, 
\footnotetext{TANG and ACTT incorrectly considered reverse bit ordering. We only consider forward bit ordering for these two methods.}
READ \cite{marchetti2019read},
ACTT \cite{verma2018actt}\footnotemark[\value{footnote}]\textsuperscript{,}\footnote{For the $R^2$ threshold, we used $0.4$.  For the 5/10 logs tested that contained no diagnostic packets; this method is equivalent to the baseline. 
},
LibreCAN (Phase 0) \cite{Pese2019librecan}\footnote{
    The authors state that the optimal choice for parameter $T_{p0,2}$ (percent decrease of bit flip rates) was between $.01$ and $.02$ depending on the vehicle. The authors likely meant between $.1$ and $.2$, because a threshold of 1\% or 2\% would lead to (and we verified this)  a \textit{very} high false positive rate. 
    For the results reported,  $T_{p0,2} = .2$ was used, resulting in much higher F-scores. 
}. 
See Section \ref{sec:related-works} for a description of each algorithm. Note that we did not test the algorithm proposed by Markowitz and Wool \cite{markovitz2017field} because it was tested by READ and shown to produce far inferior results. 
We also tested against a baseline method that simply uses constant bits as signal boundaries and assumes big endian, unsigned encodings. 
This represents accuracy scores obtained by simply identifying the obvious boundaries. 


In this section, we quantitatively compare the signal boundary classifiers (Step 1) proposed by each method (Section \ref{sec:eval-boundary-quant}, Table \ref{tb:compare} (top)), as well as the full tokenization and translation (Step 1--3) efforts of each (Section \ref{sec:eval-l1-error}, Table \ref{tb:compare} (bottom)). 
Recall that Step 2, Endianness Optimization, relies on boundary probabilities computed via Step 1 under both possible endiannesses and that Steps 1 and 2 together result in tokenization. 
Because the results of Step 2 are inherently tied to Step 1, we do not offer a stand-alone evaluation of our solution for Step 2. 
We also do not provide any comparative evaluation of Step 2 or 3 because no other methods propose algorithms for endianness optimization or signedness classification, although we did provide the classification metrics for our signedesss classifier (Step 3) in Section \ref{sec:signed_class}. 
Moreover, we do not quantitatively evaluate the interpretation (Step 4) efforts by ACTT, LibreCAN, or CAN-D because, as pointed out by Pes\'{e} et al. \cite{Pese2019librecan}, ground truth interpretations are highly subjective and difficult to evaluate quantitatively.  
Instead, we offer a qualitative comparison of the full decoding efforts in Section \ref{sec:qual_results} and Fig. \ref{fig:qual_compare_paylods_signals}, which includes the supplemental interpretations given by CAN-D. We note that ACTT's interpretation is virtually identical to CAN-D's, and LibreCAN's requires an extra tool to obtain body-related labeled time series, so we did not attempt to perform their interpretation method.

Finally, we note that READ and LibreCAN make efforts to categorize signals, which is an added benefit of these methods over ours, but we do not evaluate the efficacy of their categorization algorithms.

\subsection{Signal Boundary Classification Evaluation}
\label{sec:eval-boundary-quant}
We first quantitatively evaluated the signal boundary classification algorithms of each method using  three test sets that differ in the number of positive labels (detailed in Table \ref{tb:test_set_pos_labels}).

\begin{wraptable}[9]{r}{2.7cm}
 \fbox{%
   \begin{minipage}{\dimexpr2.7cm-2\fboxsep}
    \small
    \caption{Positive labels in each test set (5784 negative labels in all sets)}
    \label{tb:test_set_pos_labels}
        \begin{tabular}{l r r}
        \toprule
        {} &   {$n$} &  \% \\
        \midrule
        c  &   834 & 13\\
        f$-$ &   208&  3\\
        f$+$ &  1159 & 17 \\
        \end{tabular}
           \end{minipage}}
\end{wraptable} 
The condensed ``c'' set uses all positive labels (boundaries) in condensed traces (constant bits removed), thus increasing the number of nonobvious positive labels and decreasing class bias, resulting in the most robust evaluation set for testing and comparing the efficacy of signal boundary classification algorithms. 
In the full (noncondensed) ``f$+$'' set, all nonconstant samples are scored (including obvious examples of LSBs abutting constant bits/message ends). 
These full (noncondensed) f$+$ traces give a more accurate representation of the (very biased) distribution of labels and the most realistic setting for the positive and negative samples. 
This f$+$ set is the most representative and will yield the most realistic metrics for the total signals that could be extracted using a given method. 
Finally, in the full nonobvious set ``f$-$'', only nonobvious examples (those not abutting constant bits) are scored. This test set has very few positive labels (3\%), but unlike c, all are boundaries that delimit two adjacent signals in actual data, and unlike f$+$ will not result in score inflation from obvious boundaries not attributable to the algorithm being scored. 
The f$-$ set gives a balance of realism in use without the inflation of metrics from the obvious boundaries. 

To illustrate the differences in the three sets, suppose for the moment that bit 3 is the only constant bit observed in an ID's 64-bit data field; thus, bit 2 is necessarily the LSB of a signal, and bit 4 is the MSB of a signal. 
In the condensed testing set, c,  bit 3 is removed so bit 2 (LSB) and bit 4 (MSB) abut,  allowing  a positive example (i.e., a signal boundary that the algorithm will have to decide). Thus, c increases the ratio of positive examples in the data and allows more robust evaluation of whether the algorithm can identify signal boundaries. 
In this same case, the f$+$ set and the f$-$ set simply leave the constant bit (bit 3) in the dataset, allowing immediate and obvious identification of a signal boundary. 
The difference is that f$+$ indicates that this obvious signal boundary is counted during testing, whereas f$-$ indicates that it is ignored in testing because it is obvious (i.e., does not require inference based on neighboring bits). 
Thus, f$+$ scores indicate efficacy when used on real data, although the many obvious signal boundaries mask the performance of the classifiers. Sets c and f$-$ are included to provide this insight.

\begin{table}[!b]
    \vspace{.2cm}
   \centering
    \begin{threeparttable}
\begin{tabular}{lccccccccc}
    & &  & \rotatebox{80}{\textbf{Baseline}} &  \rotatebox{80}{\textbf{TANG} \cite{nolan2018unsupervised}} &  \rotatebox{80}{\textbf{ACTT}\cite{verma2018actt}} & \rotatebox{80}{\textbf{READ} \cite{marchetti2019read}} &  \rotatebox{80}{\textbf{LibreCAN} \cite{Pese2019librecan} } & \rotatebox{80}{\textbf{CAN-D Heuristic}}& \rotatebox{80}{\textbf{CAN-D ML}}\\
\toprule
    \multirow{9}{*}{\rotatebox{90}{\textbf{Signal Boundary Classif.}}}
        &\multirow{3}{*}{\textbf{F}}
            &c     & 0.0 &  67.1 &   *      &  71.1 &      78.7 &            91.6 &     \textbf{93.5} \\
            &&f$-$ & 0.0 &  45.6 &  14.5    &  63.2 &      70.6 &            89.6 &     \textbf{91.2} \\
            &&f$+$ & 90.1&  84.9 &  89.6    &  94.4 &      94.2 &            98.1 &     \textbf{98.4} \\

    \cdashlinelr{2-10}
        &\multirow{3}{*}{\textbf{P}}
            &c     & 0.0   &  62.6     &    *    &  94.8 &      86.8 &            \textbf{97.2} &     96.4 \\
            &&f$-$ & 0.0   &  31.8     & 35.2    &  80.6 &      64.2 &            \textbf{89.9} &     87.6 \\
            &&f$+$ &\textbf{100.0} &  75.7 &  96.5     &  97.6 &      92.4 &            98.2 &     97.6 \\
    \cdashlinelr{2-10}
        &\multirow{3}{*}{\textbf{R}}
            &c     &   0.0 &  72.4 &   *        &  56.8 &      71.9 &            86.7 &     \textbf{90.8} \\
            &&f$-$ &   0.0 &  80.3 &    9.1   &  51.9 &      78.4 &            89.4 &     \textbf{95.2} \\
            &&f$+$ &   82.0&  96.5 &    83.7   &  91.4 &      96.1 &            98.1 &     \textbf{99.1} \\
\midrule \\
\midrule 
\multirow{11}{*}{\rotatebox{90}{\textbf{Mean $\ell^1$ Signal Error }}}
& \multirow{10}{*}{\rotatebox{90}{CAN  Log}}
    &\#1   &      18.9 &  30.0 & 18.9  &  20.9 &      25.4 &             4.0 &      \textbf{3.9} \\
    &&\#2     &      11.2 &  16.3 & 12.7  &   9.7 &      11.1 &             \textbf{0.7} &      1.9 \\
    &&\#3    &      20.9 &  25.6 & 20.9 &  19.8 &      20.5 &             \textbf{1.3} &      2.1 \\
    &&\#4    &       4.4 &  13.8 & 6.4  &   2.3 &       3.5 &             \textbf{1.3} &      1.7 \\
    &&\#5    &      11.9 &   8.9 & 11.9  &   8.8 &       3.7 &             3.3 &      \textbf{2.3} \\
    &&\#6    &      11.0 &  18.4 & 11.0  &  10.6 &       9.7 &             4.9 &      \textbf{3.4} \\
    &&\#7    &       6.1 &  16.4 & 7.9  &   5.5 &       7.6 &             \textbf{0.0} &      \textbf{0.0} \\
    &&\#8    &       9.4 &  16.7 & 9.4  &   7.9 &      10.2 &             \textbf{0.2} &      1.5 \\
    &&\#9    &      11.1 &  14.2 & 10.6  &  12.0 &      12.8 &             1.5 &      \textbf{1.2} \\
    &&\#10    &       9.8 &  14.1 & 11.3 &   7.9 &       9.9 &             \textbf{0.7} &      0.9 \\
    \cdashlinelr{2-9}
    &\multicolumn{2}{c}{Average} & 11.5 &  17.4 &  11.9 &  10.5 &      11.4 &  \textbf{1.8} &      1.9\\
\bottomrule
\end{tabular}
\caption{\textbf{Top:} Comparison of signal boundary classification results presented. \textbf{F} = F-score, \textbf{P} = precision, \textbf{R} = recall. 
We test each method using the three test scenarios, denoted in the third column and described in Sec. \ref{sec:eval-boundary-quant}.
``Baseline'' identifies only obvious signal boundaries only at constant bits, which trivially has perfect precision.  
CAN-D ML achieves the highest F-score and recall, while the heuristic exhibits the best precision for all sets.  
Both exhibit over $\smalltilde$10\% improvement in recall over all previous methods in the two  difficult test sets (c, f$-$).
We do not evaluate ACTT under scenario (c); it relies heavily on constant bits to shrink the search space. 
\textbf{Bottom:} Mean $\ell^1$ error of translated signal values (Eq. \ref{eq:log_error}) reported for each CAN log, \#1--\#10 (described in Table \ref{tb:log_signal_stats}). 
CAN-D has lowest error on every log (often by far), exhibiting more than five times lower error on average, and almost five times times lower error on the ``gold-standard'' log \#1.
Averaged across all CAN logs, CAN-D is the only algorithm to substantially beat the baseline,
achieving perfect translation in log \#7.
The most substantial decrease in error is on logs that have a high percentage of little endian (e.g., \#1, \#9) or signed (e.g., \#2, \#3) signals, demonstrating the efficacy of Steps 2 and 3.
Even on \#4, the lone log without little endian or signed signals, CAN-D exhibits at least $\smalltilde$ 50\% decrease in error from other methods.
Finally, note that while CAN-D ML has slightly higher average error than CAN-D heuristic (due mostly to worse precision in Step 1), it has lower error for all logs containing little endian signals (\#1, \#6, \#9), perhaps illustrating that the endianness optimization (Step 2) benefits from the probability inputs offered by the ML in Step 1. 
}
\label{tb:compare}
\end{threeparttable} 
\end{table}

The classification F-score, precision, and recall for each scenario are reported in Table \ref{tb:compare} (top). Recall that because little endian signals are split on the byte boundary into two big endian signals for labeling, we test only the efficacy of the signal boundary classification methods without taking endianness into account, 
thus not penalizing other algorithms for the limiting assumption of big endianness. Also note that because CAN-D is supervised, reported metrics are from aggregating results from LOOCV per log.

\subsection{Signal Error Evaluation}
\label{sec:eval-l1-error}
Second, we compare the full tokenization and translation efforts of each method, computing the $\ell^1$ error between the translated signals and their corresponding ground truth signals. 
Results are shown in Table \ref{tb:compare} (bottom).  
The motivation for this evaluation is that ultimately, the goal of all of these methods to extract time series that can be used as actual real-time measurements from systems in the car. 
Therefore,  the most important metric for measuring the efficacy of these methods is not how many bits overlap or the number of boundaries correctly classified (as described above), but the difference between the values of the extracted signal's time series and the true signal's time series. 
All previous methods assume big endian, unsigned signals; consequently, once signal boundaries are assigned, the translated signal values are completely determined, and this is what is used for this second evaluation. For CAN-D, Steps 2--3, Endianness Optimization and Signedness Classification, provide the remaining tokenization and translation information. 

We compute the score for each log as follows. 
Let $S$ denote the set of normalized true signals and $\hat{S}$ the set of normalized predicted signals (all taking values in [0,1])  for a CAN log. 
Let $\eta:S\to\hat{S}$ so that for each true signal, $s$, $\eta(s)$ is the predicted signal that contains the MSB of $s$. 
Any predicted signals that are left unmatched ($\forall \hat{s}\in\hat{S}\backslash \eta(S)$) are paired with the zero vector $\vec{0}$. 
Take the normalized $\ell^1$ difference between each signal pair, resulting in a signal error between 0 and 1. 
The \textit{mean signal error for the log} is defined as
\begin{equation}
    \label{eq:log_error}
       \sum_{s\in S} \|s-\eta(s)\|_1 + \sum_{\hat{s}\in \hat{S}\backslash\eta(S)} \|\hat{s}\|_1 \; , 
\end{equation}
where $\|s\|_1:= \sum_{t = 1}^{n_{id}} |s(t)|/n_{id}.$

     \begin{figure*}
        \vspace{-.8cm}
    \caption{Tokenization and translation of three messages by CAN-D and top competing methods, READ \&  LibreCAN.
    When interpretation is provided by CAN-D, the label and units of the matched diagnostic are shown with the $R^2$ value, and the values are scaled appropriately. 
     }
    \hrule
    \centering
        \begin{subfigure}[c]{\textwidth}%
        \includegraphics[width=.21\textwidth]{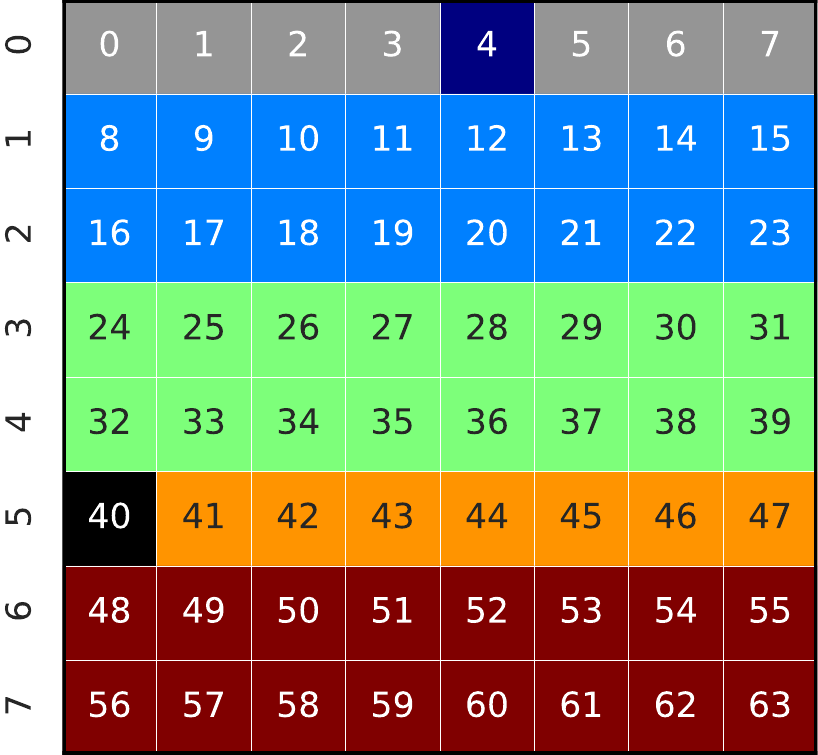}%
        \includegraphics[width=.38\textwidth]{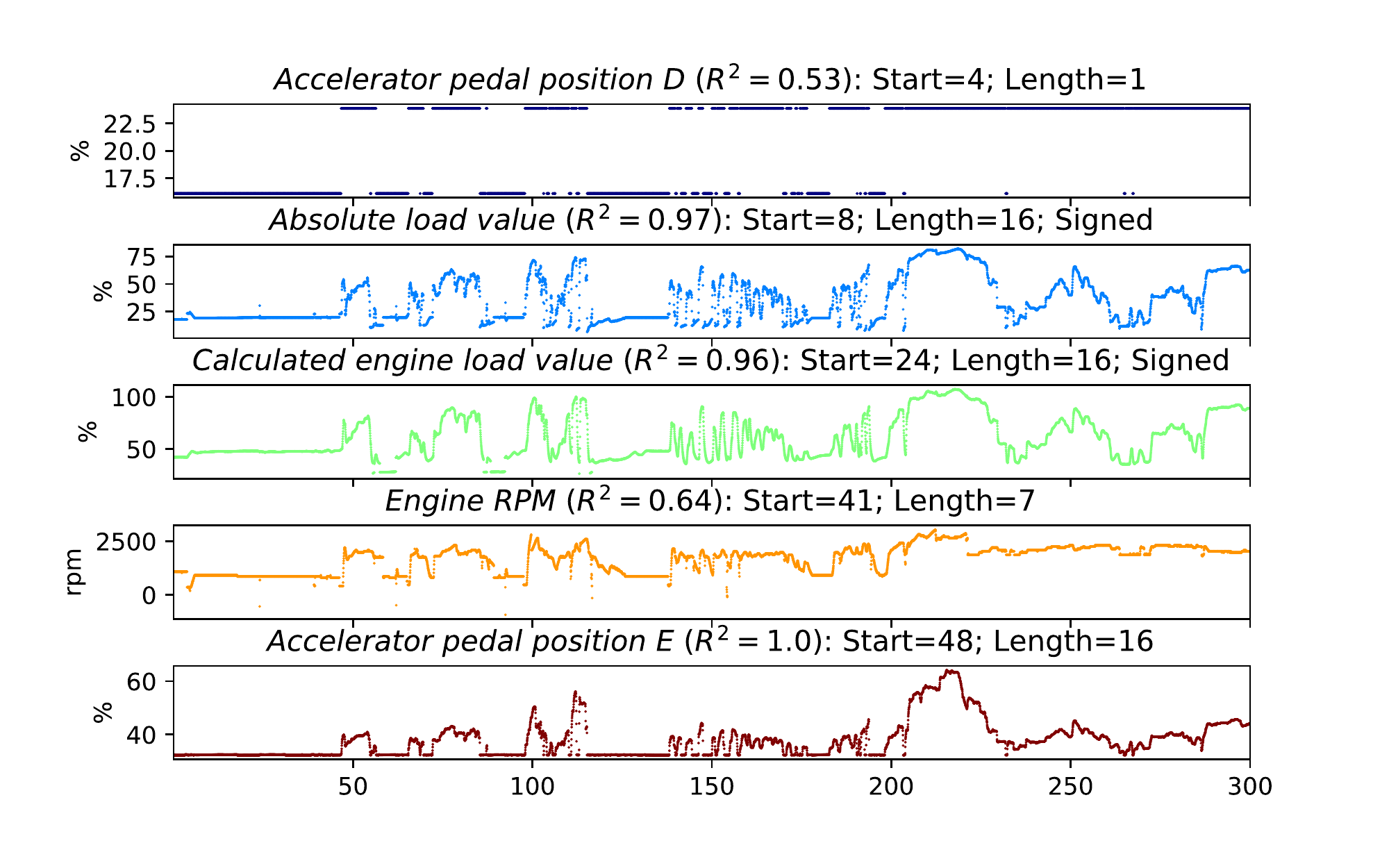}%
        \includegraphics[width=.39\textwidth]{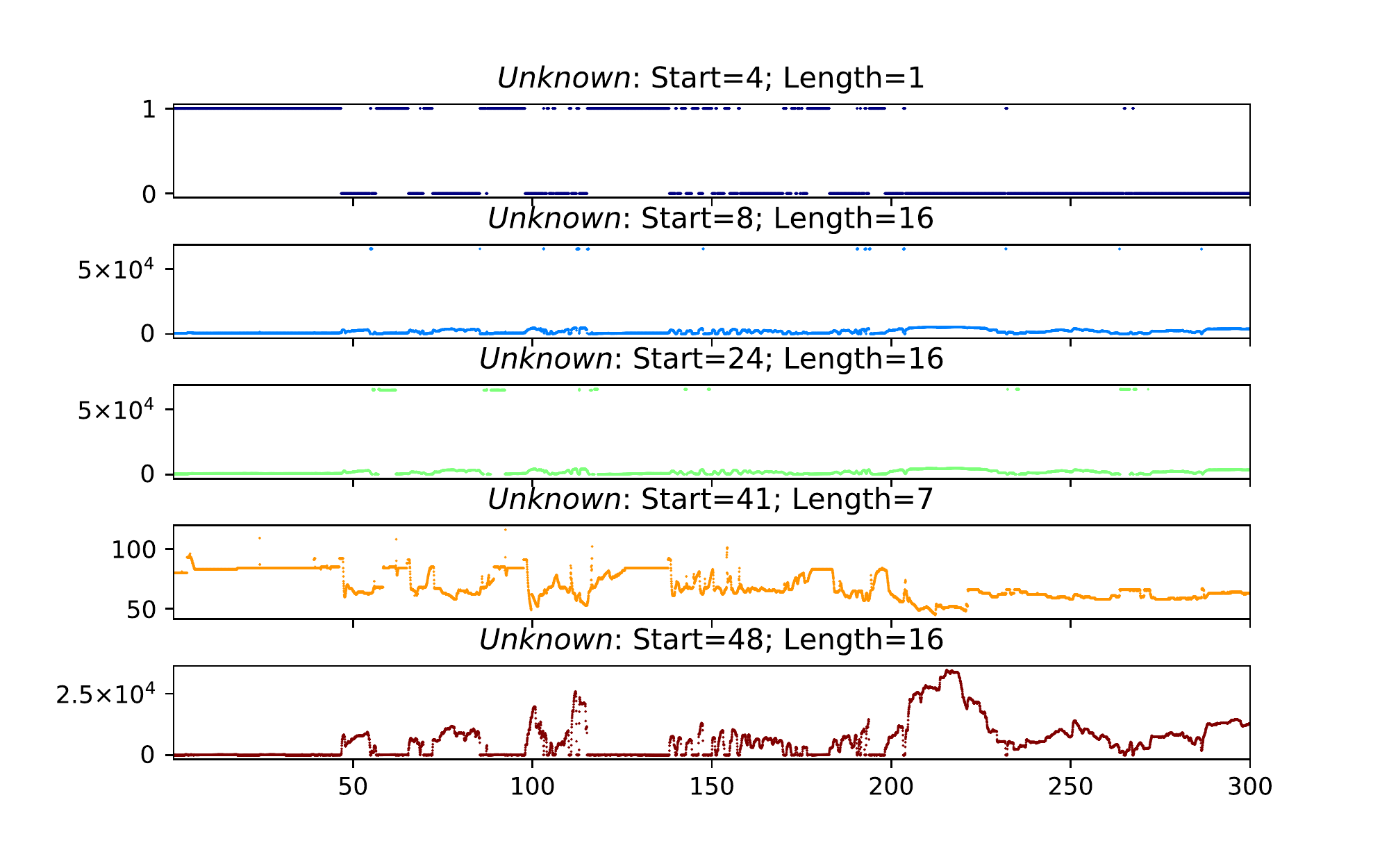}%
        \vspace{-.1cm}%
        \subcaption{Message containing signed and unsigned engine- and pedal-related signals. \textbf{Left}: Signal boundaries and endianness are correctly identified by all methods. \textbf{Middle}: All signals are correctly translated and have physical interpretations by CAN-D. Highly correlated matches found for \textcolor{green}{\textbf{green}}, \textcolor{Cerulean}{\textbf{blue}}, and \textcolor{Maroon}{\textbf{maroon}} signals. The \textcolor{Blue}{\textbf{navy}} signal at bit 4, matched to DID `Accelerator pedal position D' with low correlation ($R^2 = .53$), is likely an accelerator indicator. As this is not an available DID, CAN-D has unearthed information that could not be simply queried.
        \textbf{Right}: Other methods incorrectly translate \textcolor{green}{\textbf{green}} and \textcolor{Cerulean}{\textbf{blue}} signals as unsigned, resulting in sharp discontinuities where the signals change sign.%
        }%
    \end{subfigure}%
    \vspace{-.15cm}
    \hrule%
    \begin{subfigure}[c]{0.485\textwidth}%
        \includegraphics[width=.99\textwidth]{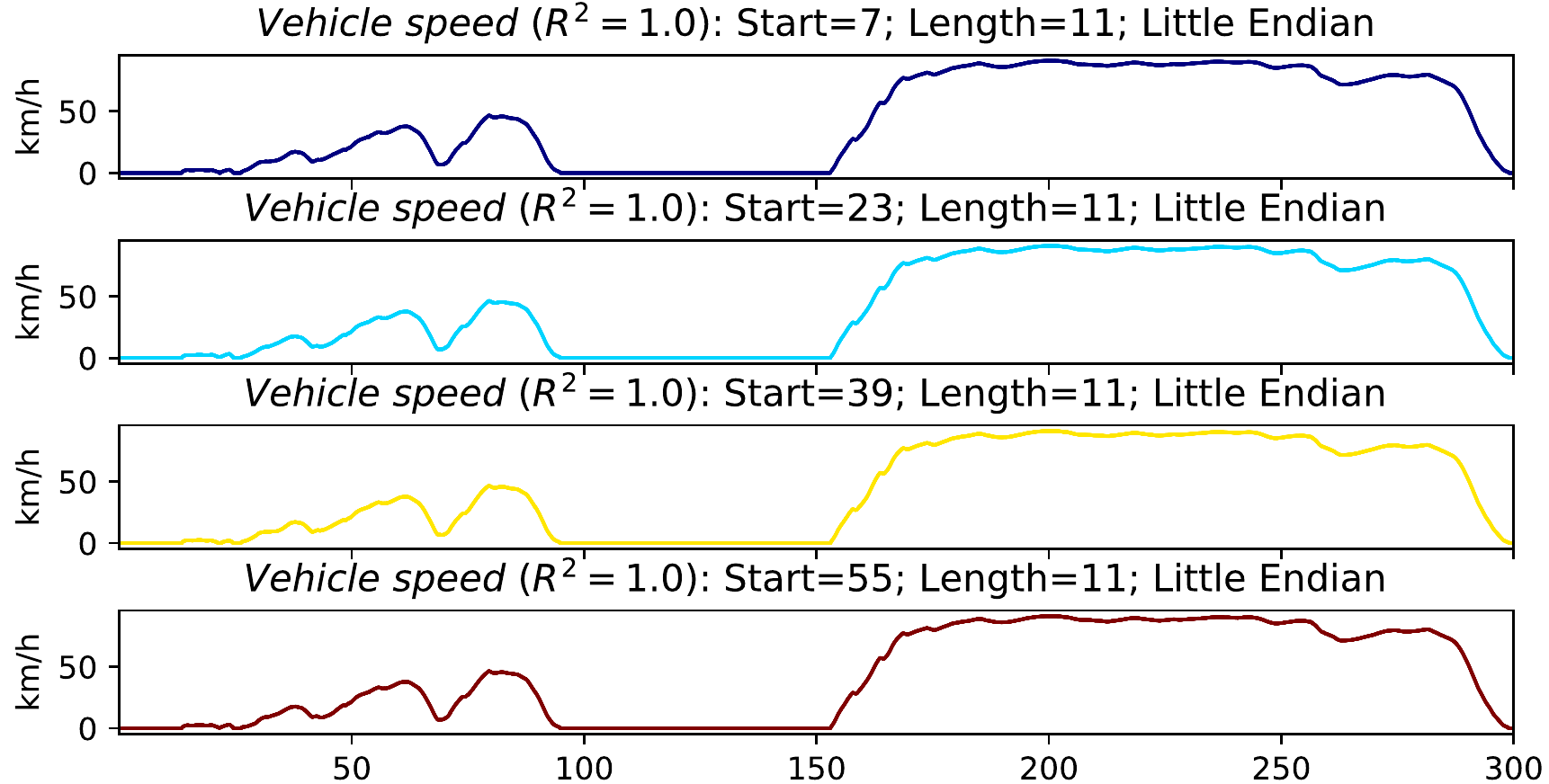}
        \raggedleft{\rule[.05mm]{4.4cm}{.01cm}} \\
        \includegraphics[width=.45\textwidth]{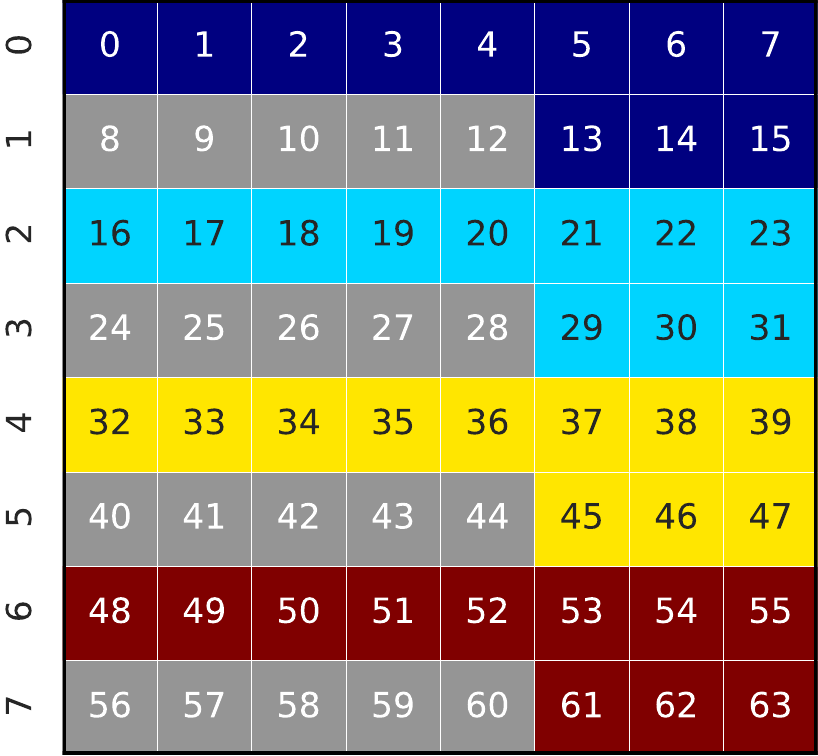}
        \vspace{-.4cm}
        \hspace{.2cm}
        \rule[-2mm]{.005cm}{4cm}
         \hspace{.2cm}
        \includegraphics[width=.45\textwidth]{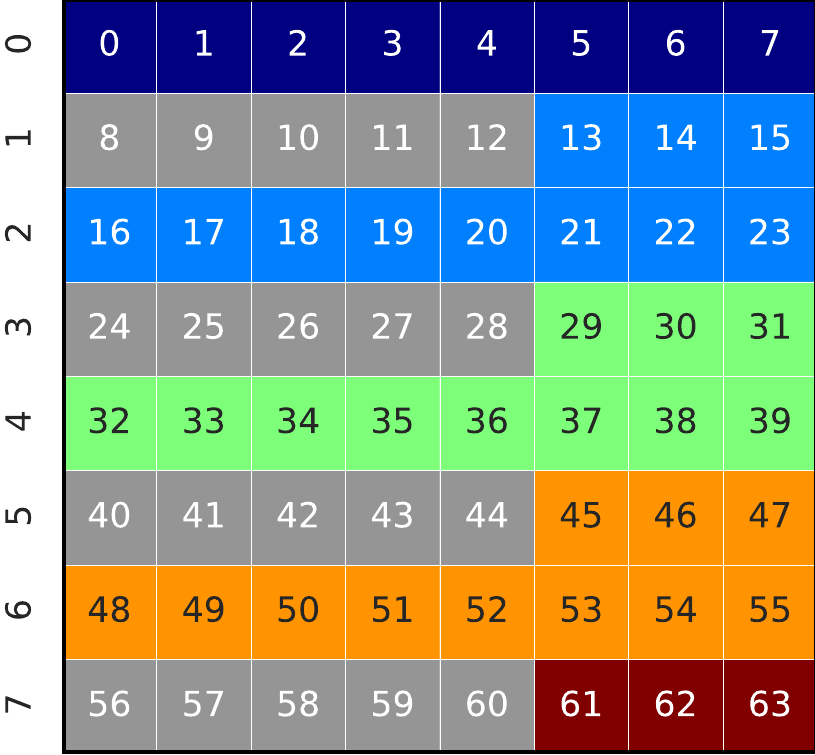} \\
        \raggedright{\rule[-2mm]{4.4cm}{.01cm}}
        \includegraphics[width=.99\textwidth]{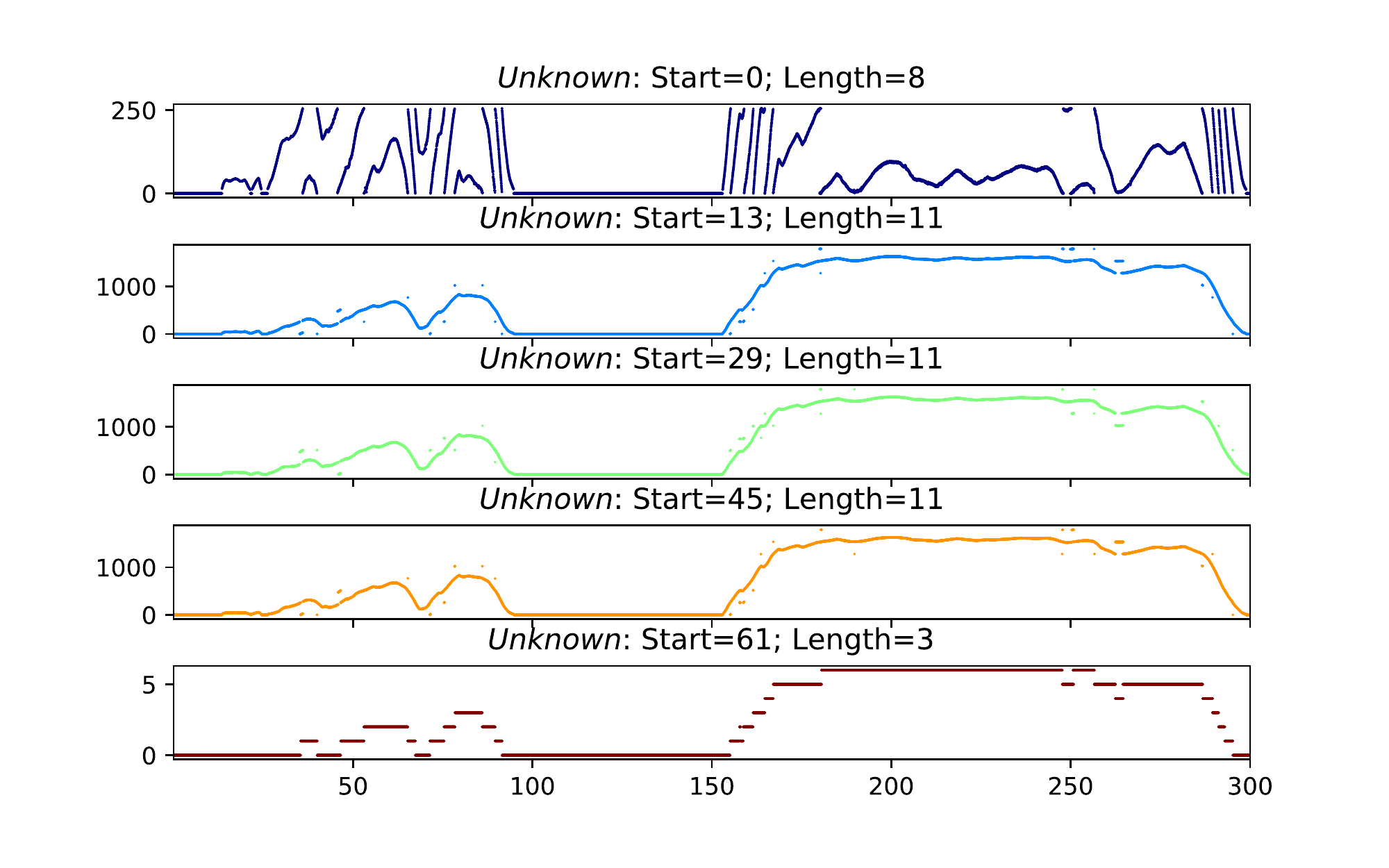}
        \vspace{-.45cm}
        \subcaption{Message containing four wheel speeds encoded as little endian signals. \textbf{Top}: Correct tokenization and translation by CAN-D and match to ``Vehicle Speed'' DID with $R^2 = 1$. \textbf{Bottom}: Mistokenized as five big endian signals by other methods with MSBs (bits 13--15, 29--31, and 45--47) attributed to the wrong signals. Because all encode speed, \textcolor{Cerulean}{\textbf{blue}}, \textcolor{green}{\textbf{green}}, and  \textcolor{Orange}{\textbf{orange}} signals appear correct, save some minor discontinuities. 
        However, these signals encode the wheel speeds and are often used by Electronic Stability Control to stimulate anti-lock braking and traction control pending discrepancies in wheel speeds; mixing the MSBs of wheel speeds may go unnoticed in normal conditions but could prove consequential in adverse driving conditions!
        }
    \end{subfigure}
    \hspace{.3cm}
    \begin{subfigure}[c]{0.485\textwidth}
        \includegraphics[width=\textwidth]{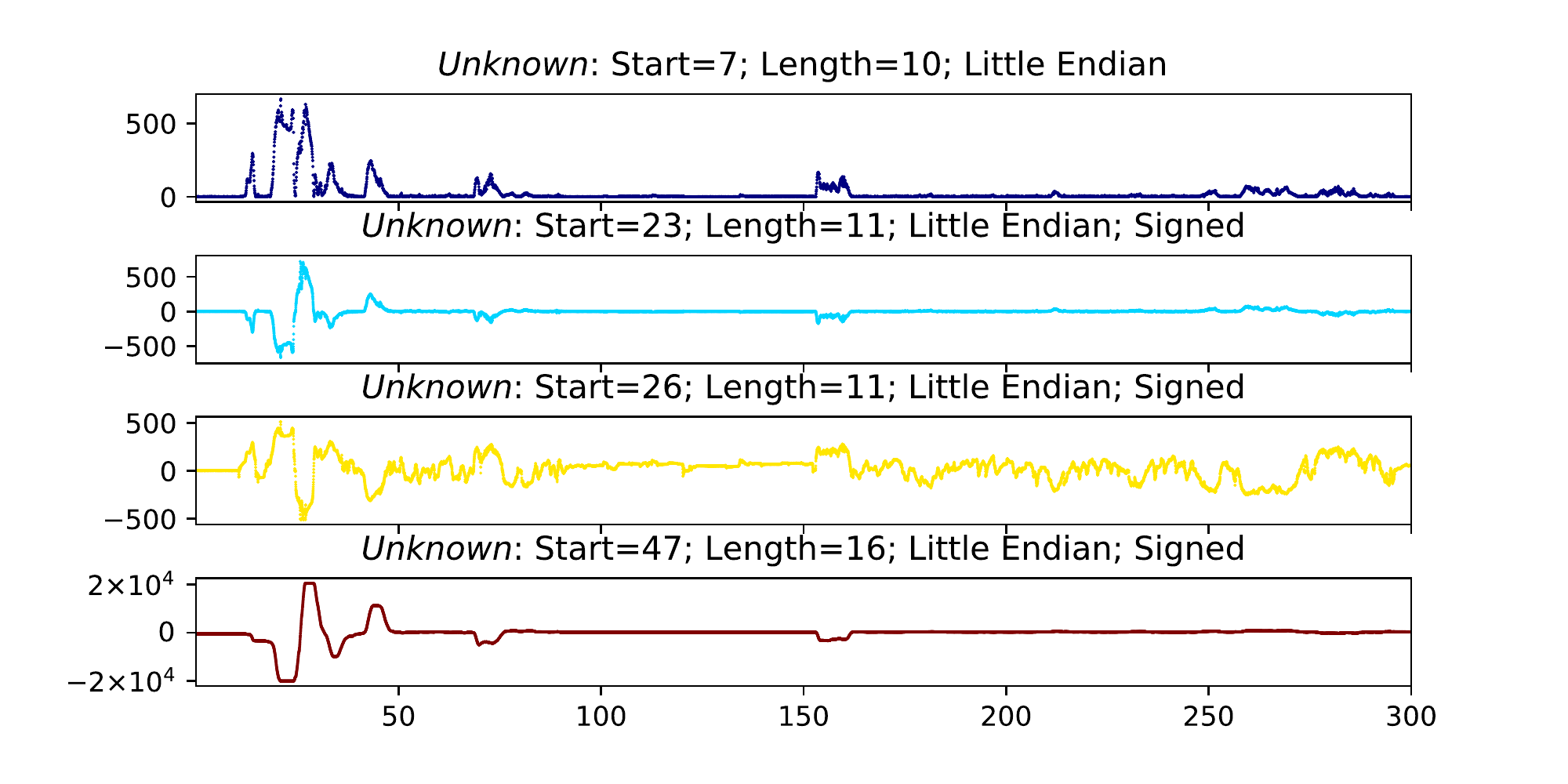}
        \raggedleft{\rule[.05mm]{4.4cm}{.01cm}} \\
        \includegraphics[width=.45\textwidth]{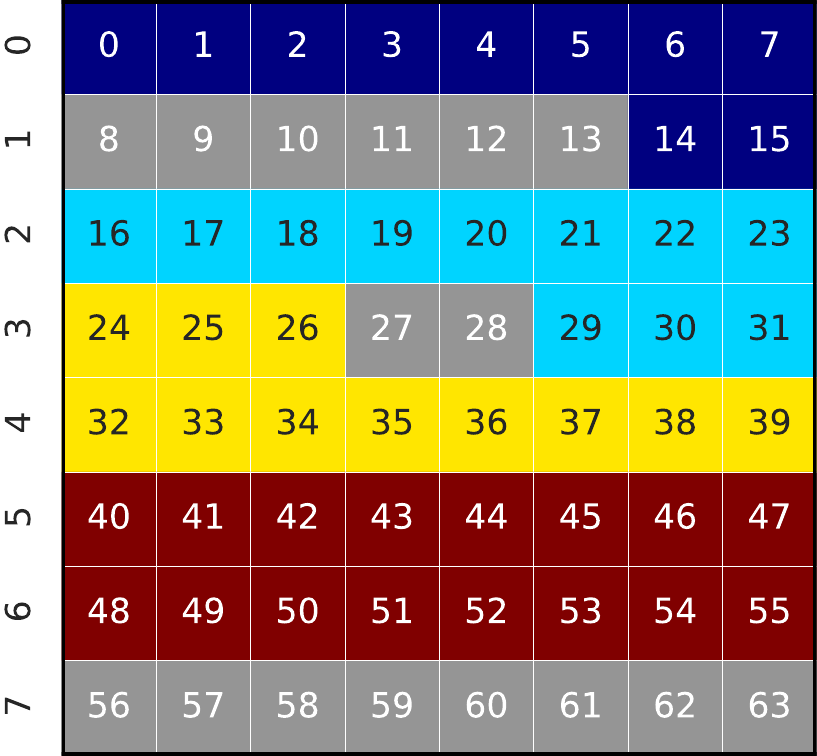}
         \vspace{-.4cm}
        \hspace{.2cm}
        \rule[-2mm]{.005cm}{4cm}
         \hspace{.2cm}
        \includegraphics[width=.45\textwidth]{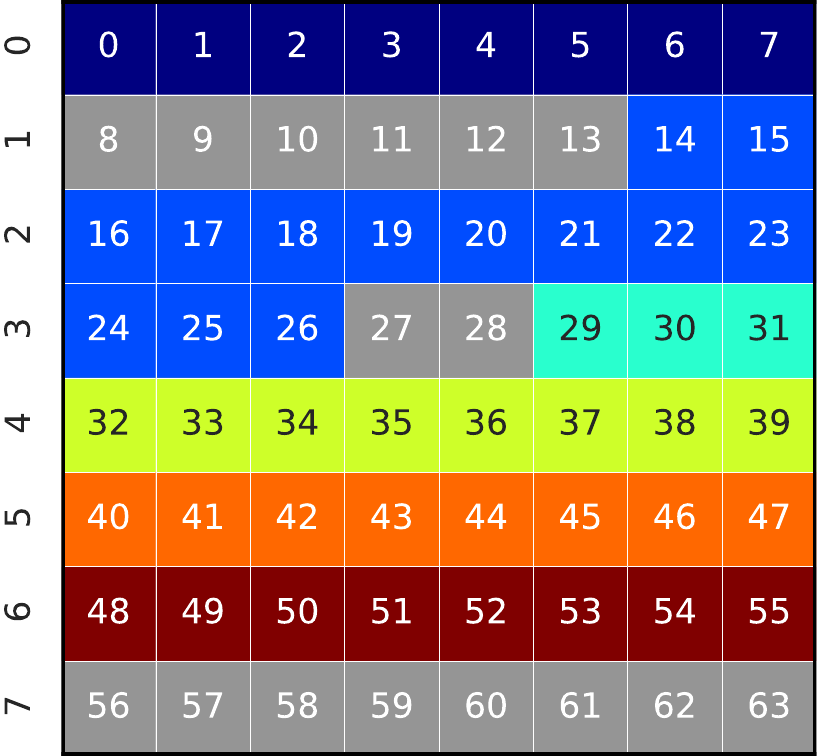} \\
         \raggedright{\rule[-2mm]{4.4cm}{.01cm}}
        \includegraphics[width=.99\textwidth]{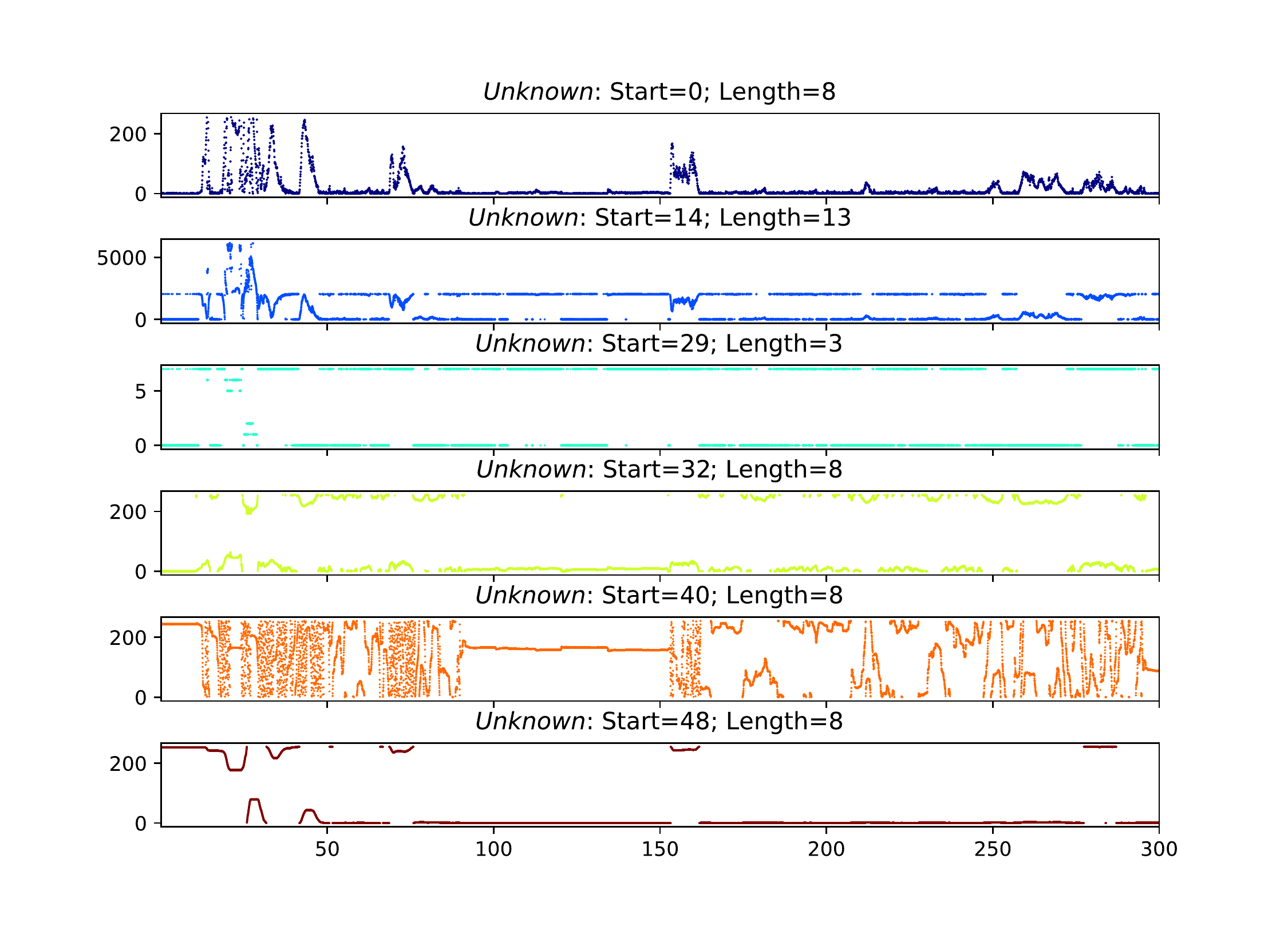}
        \vspace{-.45cm}
        \subcaption{Message containing four steering-related, little endian signals, three of which are signed. \textbf{Top}: Correct tokenization and translation by CAN-D (no interpretation). \textbf{Bottom}: Incorrect tokenization and translation by other methods. 
        Assuming big endian signals, they are forced to cut on most byte boundaries, resulting in truncated, noisy \textcolor{SeaGreen}{\textbf{teal}}, \textcolor{SpringGreen}{\textbf{lime}}, \textcolor{Orange}{\textbf{orange}}, and \textcolor{Maroon}{\textbf{maroon}} signals. 
        The \textcolor{Blue}{\textbf{navy}} signal does not appear noisy but is noticeably incorrect when comparing the scale and the values for $t \in [0,50]$ to the correct CAN-D translation. 
        The two MSBs are misattributed to the next signal, resulting in errors of at least $2^8$ when the MSB(s) are nonzero.
        }
    \end{subfigure}
     \label{fig:qual_compare_paylods_signals}
\end{figure*}

    \subsection{Qualitative Results}
\label{sec:qual_results}
Fig. \ref{fig:qual_compare_paylods_signals} depicts three examples of messages decoded by CAN-D (identical decodings for the ML and heuristic signal boundary classification) and by the most accurate competing methods (READ and LibreCAN, which both produce the same signal boundary predictions for these examples)  with detailed descriptions and discussion. 
These examples illustrate a message with 
signed and unsigned signals (top), 
little endian unsigned signals (bottom left) 
and little endian (signed and unsigned) signals (bottom right). 
CAN-D correctly tokenizes and translates all examples depicted and overall furnishes interpretable time series. 
Where available, CAN-D's physical interpretation  (Step 4, Section \ref{sec:diag-matching})  is provided in annotations above signals, showing $R^2$ value to gauge goodness-of-match. 
Overall, mistokenization and mistranslation by other methods resulted in rampant discontinuities and dramatic error in most time series, exhibiting the necessity of correctly identifying each signal's endianness and signedness. 

We also attempted to find common attributes of the incorrectly classified signals by CAN-D and found that two key characteristics that lead to higher signal error. 
First, discontinuous signals or discrete signals are often incorrectly split into multiple signals. 
Most commonly, this seems to occur for CAN signals that are mapped to enumerated choices known as ``enumeration value types'' \cite{vector_candb_docs} (e.g., a gear shift signal with the mapping $\{2:\texttt{P}, 4: \texttt{R}, 8 : \texttt{D}, 16: \texttt{L}\}$). Secondly, short (1--3 bit) abutting signals that move together are sometimes incorrectly joined together because of the fact that the bit flips of adjacent MSBs and LSBs of these signals are highly dependent. Importantly, other methods appear to have similar shortcomings. 

\section{Prototype OBD-II Plugin}
\begin{wrapfigure}[10]{r}{.25\textwidth}
    \centering
    \includegraphics[scale = .04]{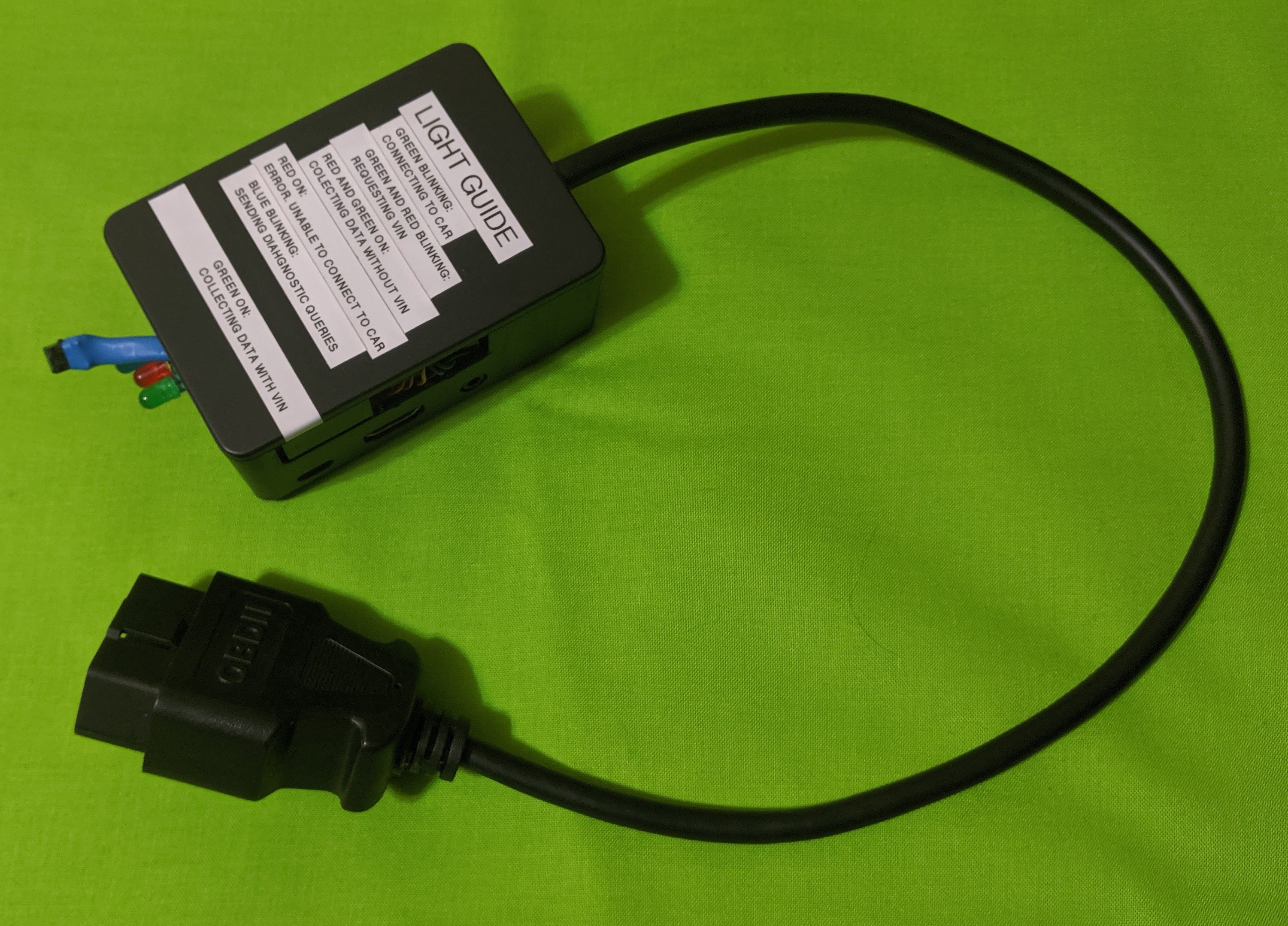}
    \caption{Prototype CAN-D device using a Raspberry Pi and CANBerry Dual 2.1 boards.}
    \label{fig:hw-pic}
\end{wrapfigure}

The CAN-D prototype device is a vehicle-agnostic, OBD-II plugin  that collects CAN data from the vehicle and runs the entire CAN-D pipeline depicted in Fig. \ref{fig:pipeline}. 
The prototype (shown in  Fig. \ref{fig:hw-pic}) is built using Linux-based, single-board computers. 
\label{sec:hw} 
\vspace{-.1cm}
Specifically, we use a Raspberry Pi 3B+ with Raspbian Buster in conjunction with an Industrial Berry's  CANBerry Dual 2.1\footnote{\url{http://www.industrialberry.com/canberrydual-v2-1/}}.
The Raspberry Pi 3B+ offers 1 GB of RAM and a 1.4 GHz ARMv8 processor. 
The device is powered either from battery or using on-board power from a vehicle's 12-volt system.


One challenge of building a vehicle-agnostic prototype is that the bitrate for the CAN is unknown and variable per  vehicle, and improper bitrate selection can cause adverse vehicle function.
To solve this issue, the device iterates through common bitrates, identifying the bitrate that results in only expected packets. 
This allows our prototype to be compatible with most CANs regardless of bitrate.

Another complication is that automobiles typically have multiple CAN buses, and often more than one is available from the OBD-II interface. 
The prototype analyzes two unique networks by allocating a dedicated CAN controller for each using CANBerry Dual 2.1. 
Once connected, it automatically determines each network bitrate and identifies the VIN using the Unified Diagnostic Services query--response protocol (see Sec. \ref{sec:can-basics}), and begins logging data. 
A switch on the prototype can be used to augment the network traffic with the other available diagnostic queries.  
The CAN traffic is collected using SocketCAN CAN-Utils\footnote{\url{https://github.com/linux-can/can-utils}}, logged to an ASCII-encoded text CAN log file, and 
named using the  VIN (automatically identified via a DID query) and timestamp. 
Flashing LEDs on the device indicate the progress throughout this collection process. 
The device then runs this CAN log through the CAN-D pipeline, outputting a DBC, which can then be used for real-time decoding and visualization of signals on the device.




Using our heuristic signal boundary classifier (Step 1), we benchmarked the device running the CAN-D pipeline.   
We collected CAN traffic augmented with diagnostic data from a passenger vehicle for 70 s, logging $\smalltilde$170K frames. 
Running the pipeline on this log averaged 129 s--55 s for preprocessing, 14 s for tokenization and translation (Steps 1--3), 50 s for interpretation (Step 4), and 10 s for writing to DBC, over six runs with negligible variance.   
As implementation was focused on algorithmic research, efficiency was not a main goal and could be significantly improved. 
This proves the pipeline is viable for use in a portable, lightweight, edge computing device. 

Finally, we note a limitation with respect to the prototype. As stated previously, vehicles not required to follow the diagnostic standards (e.g., nonemissions producing vehicles) may not be queryable, and thus getting labeled diagnostic ground truth signals for Step 4 would not be possible. On the other hand, some vehicles (particularly newer models) have segmented CANs, allowing diagnostic queries to be sent/received at the OBD-II port but not allowing visibility to ambient CAN traffic. Thus, this prototype would be unable to collect the raw CAN data via this port. One possible workaround for these vehicles would be to bypass the gateway at the OBD-II connector and directly access the targeted network's CAN wires. 



\section{Conclusion}
\label{sec:conclusion}
We considered the problem of developing a vehicle-agnostic method for extracting the hidden signals in automotive CAN data and present a comprehensive survey of this area. 
We presented CAN-D, a four-step, modular pipeline using a combination of ML, a novel optimization process, and heuristics  to identify and correctly translate signals in CAN data to their numerical time series. 
In particular, CAN-D is designed to extract big and little endian signals as well as signed and unsigned signals.  
Although this greatly enhances the complexity of the problem, these are necessary accommodations as specified by standard signal definitions. 
As our results show,  when endianness and signedness are ignored, the resulting translations are incorrect and overly noisy.  
In evaluation on 10 diverse vehicles' data, we compared CAN-D to the four state-of-the-art methods, providing a comparative study of previous methods on a more comprehensive dataset than ever previously used. 
We achieved less than 20\% of the average error of other methods and established that CAN-D is the only method that can handle any standard CAN signal.
Finally, we presented a lightweight hardware implementation for using CAN-D in situ via an OBD-II connection to first learn a vehicle's signals, and in future drives convert raw CAN data
to multivariate time series in real time. 
Because CAN signals provide a rich source of real-time data that is currently unrealized, we hope this contribution will facilitate many vehicle technology developments.

\section*{Acknowledgments}
Special thanks to Bill Kay for his helpful comments. Research was sponsored by the Laboratory Directed Research and Development Program of Oak Ridge National Laboratory, managed by UT-Battelle LLC, for the US Department of Energy (DOE) and by the DOE, Office of Science, Office of Workforce Development for Teachers and Scientists (WDTS) under the Scientific Undergraduate Laboratory Internship program.


\vspace{-.1cm}
\small
\bibliographystyle{IEEEtranN}
\bibliography{refs}

\vskip -2\baselineskip plus -1fil

\begin{IEEEbiography}[{\includegraphics[width=1in,height=1.25in,clip,keepaspectratio]{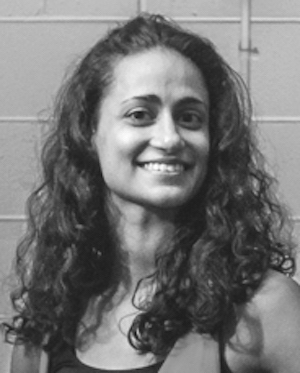}}]{Miki E. Verma} recieved a B.S. in Applied and Computational Math Sciences from the University of Washington in 2018 and is currently pursuing an M.S. in Symbolic Systems at Stanford University. Miki joined Oak Ridge National Laboratory as an intern in 2018, and is currently a research data scientist in the Cyber Resilience and Intelligence Division, where she works on applications in a variety of cybersecurity domains including vehicle and network security, and security operations. 
\end{IEEEbiography}

\vskip -2\baselineskip plus -1fil

\begin{IEEEbiography}[{\includegraphics[width=1in,height=1.25in,clip,keepaspectratio]{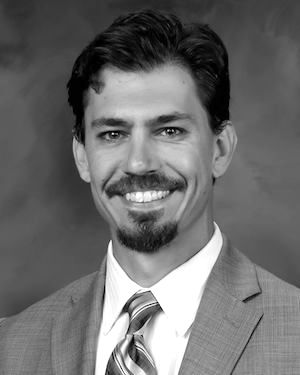}}]{Robert A. (Bobby) Bridges}  received  the B.S. in 2005 from Creighton University and a Ph.D.  in 2012 from Purdue University, both in Mathematics. Bobby joined Oak Ridge National Laboratory  as a postdoc in 2012, was promoted to Associate Researcher in 2013 then Staff Researcher in 2017, and now serves as the (Acting) Cybersecurity Research Group Leader. Bobby's provides applied mathematics and data science support to a diverse team focusing on network and vehicle security. 
\end{IEEEbiography}

\vskip -2\baselineskip plus -1fil

\begin{IEEEbiography}[{\includegraphics[width=1in,height=1.25in,clip,keepaspectratio]{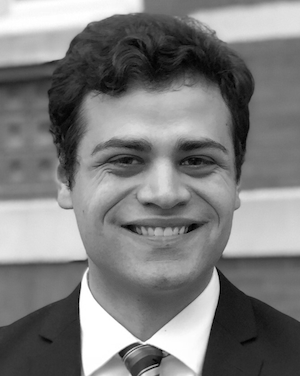}}]{Jordan J. Sosnowski} received a B.S. in Software Engineering and M.S. in Cybersecurity Engineering from Auburn University in 2019 and 2021, respectfully. Jordan joined Oak Ridge National Laboratory as a summer intern in 2019 to research controller area networks. He is currently an incident responder at a department of energy national laboratory, where his focus is on malware analysis and reverse engineering.
\end{IEEEbiography}

\vskip -2\baselineskip plus -1fil

\begin{IEEEbiography}[{\includegraphics[width=1in,height=1.25in,clip,keepaspectratio]{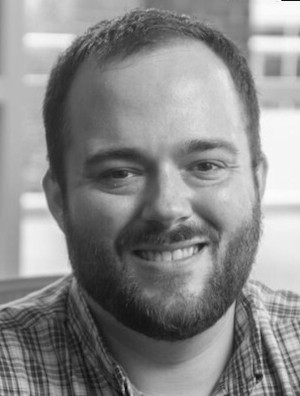}}]{Samuel C. Hollifield} is currently pursuing a B.S. degree in computer engineering from Tennessee Technological University, Cookeville, TN.

In 2018, he joined the Oak Ridge National Laboratory (ORNL) as an intern, exploring the security of electronic control units used in modern automobiles. He is currently a cyber security research associate in the Cyber Resilience and Intelligence Division at ORNL where he works primarily on network and application cybersecurity for automotives and similar embedded systems. Mr. Hollifield is a member of the Society of Automotive Engineers (SAE). 
\end{IEEEbiography}

\vskip -2\baselineskip plus -1fil

\begin{IEEEbiography}[{\includegraphics[width=1in,height=1.25in,clip,keepaspectratio]{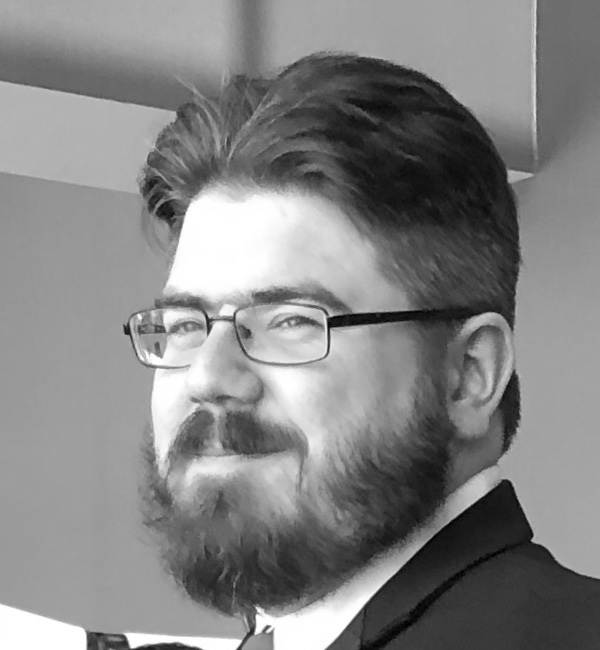}}]{Michael D. Iannacone} 
received a B.S. in computer engineering from 
Rochester Institute of Technology in 2008 and a M.S. in 
information security from Georgia Institute of Technology in 2013.

From 2007 to 2008, he was a Research Assistant with RIT's Lab for Wireless Networking and Security, modeling wireless ad-hoc networks.
From 2009 to 2011, he was a Research Assistant at the Georgia Tech Information Security Center (GTISC) research lab, studying cellular device security.
Since 2011, he has been a researcher with Oak Ridge National Laboratory, currently in the Cyber Resilience and Intelligence Division. His work has focused on network intrusion detection and anomaly detection, and securing embedded devices, vehicles, and vehicle infrastructure.

\end{IEEEbiography}









\end{document}